# IMAGING AT THE MESOSCALE (LEEM, PEEM)

A. SALA
DEPARTMENT OF PHYSICS, UNIVERSITY OF TRIESTE
PIAZZALE EUROPA 1
34127 TRIESTE, ITALY
ISTITUTO OFFICINA DEI MATERIALI – CNR
S.S. 14 KM 163.5 IN AREA SCIENCE PARK
34149 TRIESTE, ITALY

TABLE OF CONTENTS

**LIST OF ACRONYMS**

| | |
|---|---|
| AEM | Auger Electron Microscopy |
| ARPES | Angle-Resolved PhotoElectron Spectroscopy |
| CTF | Contrast Transfer Function |
| EELM | Electron Energy Loss Microscopy |
| EELS | Electron Energy Loss Spectroscopy |
| FBZ | First Brillouin Zone |
| FWHM | Full Width Half Maximum |
| IMFP | Inelastic Mean Free Path |
| HDA | Hemispherical Deflector Analyzer |
| LEED | Low Energy Electron Diffraction |
| LEEM | Low Energy Electron Microscopy |
| MEM | Mirror Electron Microscopy |
| PEEM | PhotoElectron Emission Microscopy |
| MDC | Momentum Distribution Curve |
| MPA | Magnetic Prism Array |
| PES | Photoelectron Emission Spectroscopy |
| PSF | Point Spread Function |
| QSE | Quantum Size Effect |
| SEM | Secondary Electron Microscopy |
| SPALEED | LEED Spot Profile Analysis |
| SPELEEM | Spectroscopic PhotoEmission and LEEM |
| SPLEEM | Spin Polarized LEEM |
| TEM | Transmission Electron Microscopy |
| UHV | Ultra-High Vacuum |
| UVPEEM | UltraViolet PEEM |
| XAS | X-ray Absorption Spectroscopy |
| XMCD | X-ray Magnetic Circular Dichroism |
| XMLD | X-ray Magnetic Linear Dichroism |
| XPEEM | X-ray PEEM |
| XPD | X-ray Photoelectron Diffraction |
| XPS | X-ray Photoelectron Spectroscopy |

**Abstract**

The capability to display images containing chemical, magnetic and structural information and to perform spectroscopy and diffraction from a µm-sized area makes cathode lens electron microscopy one of the most used and reliable techniques to analyze surfaces at the mesoscale. Thanks to its versatility, LEEM/PEEM systems are currently employed to study model systems in the fields of nanotechnology, nanomagnetism, material science, catalysis, energy storage, thin films and 2D materials. In the following chapter, we will present a brief but complete review of this class of instruments. After an historical digression in the introducing section, we will show first the basic operating principles of a simple setup and then the elements that can be added to improve the performances. Later, two sections will be dedicated to LEEM and PEEM respectively. In both cases, a theoretical discussion on the contrast mechanisms will prelude to a showcase of the operating modes of the instrument, with clear examples that will show the best performances available nowadays. Finally, a brief discussion about the future developments of cathode lens electron microscopy will close the chapter.

## 1. INTRODUCTION

Cathode lens electron microscopy is a technique that uses slow electrons as information carriers [1]. Differently to the case of scanning or transition electron microscopy, in this system electrons interact with the probe at very low kinetic energy, below few hundred electronvolts (eV). In such range, the inelastic mean free path (IMFP) ensures a probing depth of just a few atomic layers, making cathode lens microscopy a surface sensitive technique. It is therefore not surprising that its history parallels the history of surface science since the early years. The first example of emission microscope with slow electrons goes back to 1932, few years after the Davisson and Germer experiment [2], when Brüche and Johansson produced the first thermoionic emission microscope [3, 4]. They used coils as magnetic lenses to produce magnified images of a hot cathod's surface on a fluorescent screen. Glass enclosures and diffusion pumps provided the vacuum system, necessary to avoid dispersion of the electrons. One year later, Brüche built the first prototype of PhotoEmission Electron Microscope (PEEM) [5], using a cold cathode illuminated by UV light (**Fig. 1a**). These milestones and the rising of theoretical electron optics gave birth to a flourishing scientific community. Several theoretical calculation concerning magnification, chromatic and spherical aberration of magnetic lenses, electrostatic mirrors and einzel lenses, were made available [6–9]. It was soon understood [10] that the resolution performances could be enhanced well above light microscopy if electrons travel through the lens system at relatively high energies - tens of keV. The suggested setup was then to place the sample on a negative bias, i.e. using it as a cathode, in order to accelerate electrons after the takeoff. This grounding principle is still used nowadays for modern microscopy. The development of electron optics suffered then a sudden stop during the 1940s, not only because of World War II, but also for technological limitations of that age.

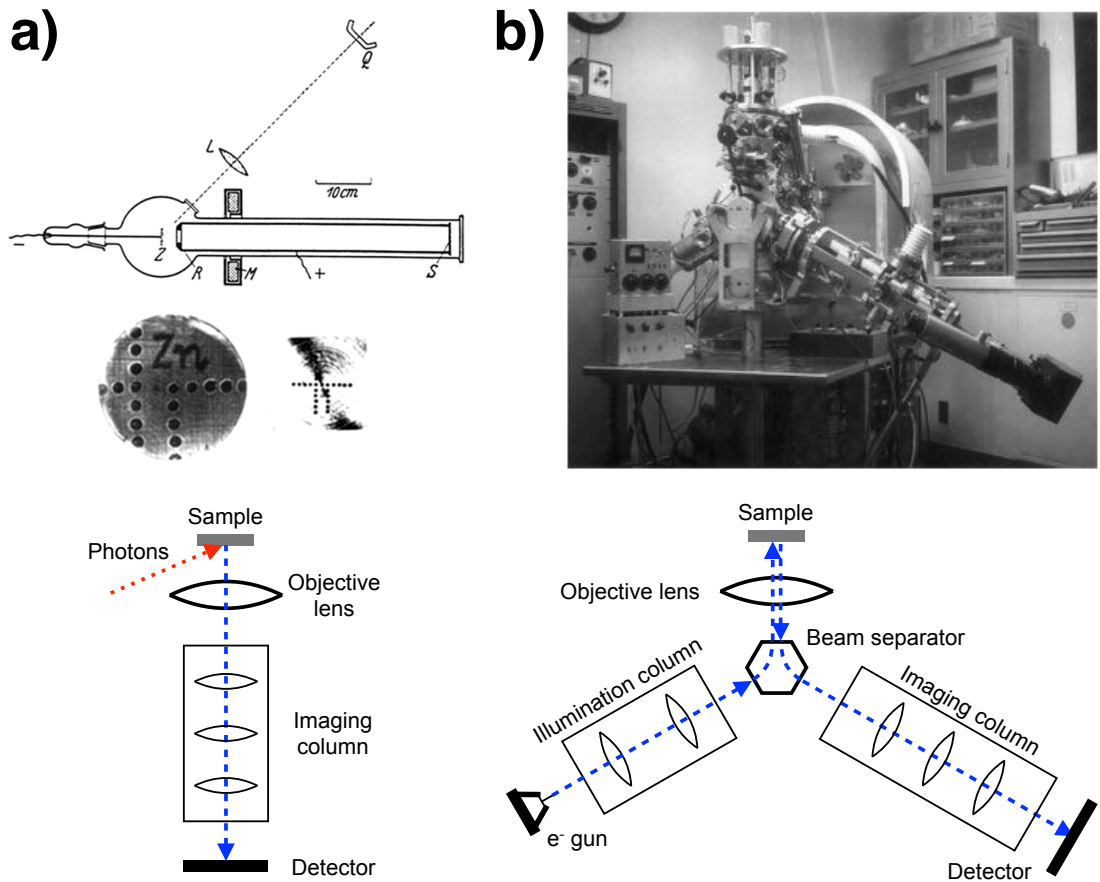

Fig. **1**: **(a)** Top: schematic of Brüche's first PEEM system and first PEEM image of scraped Zn plate with holes. Reproduced from Ref. [5] with permission, copyright 1933 Springer Science + Business Media. Bottom: scheme of an ideal PEEM system. **(b)** Top: picture of the original LEEM instrument in the 1960's. Reproduced from Ref. [11] with permission, copyright 2012 Elsevier. Bottom: scheme of an ideal LEEM system with 120° deflection.

The *renaissance* period of the 1960s coincides with the development of UHV technology: several new solutions, such as ion pumps, Cu gaskets, valves and sample transfer systems, boosted the surface science and hence the creation of more sophisticated instruments. In 1962 Ernst Bauer conceived the Low Energy Electron Microscope (LEEM) [12, 13], which uses elastically backscattered electrons as a probe (**Fig. 1b**). In this system electrons generated by a gun are decelerated to few eV before interaction with the surface. Once backscattered, the outgoing electron beam is separated from the incoming one by a magnetic field, and then processed by the lens system. In the same period, emission electron microscopy reached a period of maturity, when the demonstrated resolution of ~10 nm made explicit new limitations, such as lens aberration and astigmatism, energy dispersion, electron detection, surface stability and cleanness. The attention of the microscopy community was then gradually driven away by the success of Transmission Electron Microscopy (TEM); during the 1970s the LEEM project was frozen, while the few PEEM systems were mainly dedicated to investigate biologic system [14] and test additional light sources, such as lasers and synchrotron radiation [15]. Only in the following decade cathode lens electron microscopy manifested itself as one of the principal surface science tools. Many upgrades were introduced, enhancing the versatility and the power of such systems in the investigation of surfaces at the mesoscale. The microscopy approach was combined with other investigation methods such as spectroscopy and diffraction, already known for their successful application in surface science. In 1981 LEED patterns were displayed in LEEM systems using the backfocal plane of the objective lens [16–18], giving the possibility to spatially select the probing area. Tonner et al. demonstrated the feasibility of PEEM with synchrotron radiation in the late 1980s [19], while Ertl's group at the Fritz Haber Institute raised the interest in the chemistry community for PEEM showing the oscillatory behavior of gas adsorption on active surfaces during catalytic reactions [20]. Few years later, the first cathode lens system equipped with a hemispherical energy analyzer (Spectroscopic PhotoEmission and Low Energy Electron Microscope, SPELEEM) was planned and later hosted at BESSY and Elettra synchrotrons. The first XPEEM image, i.e. made with core-level electrons photoemitted by X-rays, was published in 1998 [21]. In the same years, two LEEM systems were equipped with spin-polarized electron gun [22, 23], pioneering magnetic imaging. Magnetic contrast was achieved also in PEEM using circular and linear dichroism of polarized light [24]. The development of new operation modes continues nowadays, with the construction of aberration-free systems towards the ultimate spatial resolution [25–27], the use of pulsed light to enable time-resolved dynamic microscopy [28], and the design of special sample holders to modify the mechanic, electric and magnetic properties of the probe *in operando* [29]. Moreover, firms started to create commercial versions of LEEM and PEEM systems. The most notable examples are Elmitec GmbH in 1995 (based on Ernst Bauer's design) and SPECS GmbH (based on Ruud Tromp's design) [30, 31].

Although the cost of such systems is relatively high compared to other microscopes, the openness of the labs to external users, e.g., at the synchrotron endstations, helped to create a vast and heterogeneous user community. Nowadays, cathode lens microscopy is widely appreciated by surface scientists. Thanks to its multidisciplinarity and to the interplay between microscopy, spectroscopy and diffraction, it has become an essential technique for the overall comprehension of surface phenomena.

## 2. Cathode lens microscopy

### 2.1 Operating principles

The first optical element electrons run into after the takeoff from the sample surface is also the most important. The cathode lens, often integrated with other refocusing elements and called objective lens, must both accelerate the electrons emitted from the sample and form a first magnified image. To do so, the sample is placed at negative potential $V$ of about 10-50 kV to act as a cathode. The first electrode is grounded and attracts the electrons towards its central aperture, where they pass through the other magnetic or electrostatic elements of the objective lens. An equipotential plot of a lens is displayed in **Fig. 2a**. The overall focal length is determined by two opposite contributions, one divergent generated by the anode aperture and one convergent generated by the other elements of the lens [32, 33]. **Figure 2b** helps to explain the relevant physics in detail.

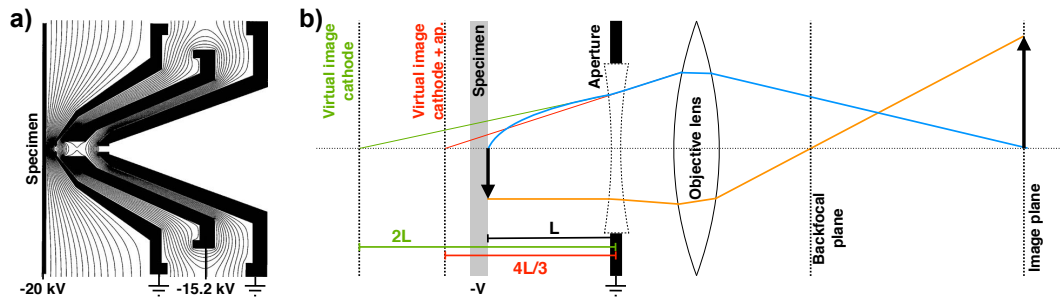

Figure 2: (a) Field contour plot of a tetrode objective lens. The equipotential lines are 1250 V apart. Reproduced from Ref. [34] with permission, copyright 2002 AIP Publishing. (b) Scheme of the cathode immersion lens. For explanation see text.

When electrons are emitted from the surface at a distance L from the entrance aperture of the microscope with kinetic energy $E_0$ and angle $\theta_0$, the acceleration in the quasi-homogeneous electric field in the cathode lens imposes a parabolic trajectory (in blue). This real situation can be converted to a virtual frame, in which electrons assume a linear trajectory and appear to be originated from a virtual image plane located at a distance 2L from the anode (green trajectory). Since in the real case the anode aperture distorts the electric field, the optical effect is to create a thin diverging lens with focal length -4L, called "aperture lens" [35]. The final virtual image is then placed at a distance 4L/3 from the anode (red trajectory), magnified by a factor 2/3. After the aperture, the field of the other elements of the objective lens magnifies the electron beam by a factor $M_M$.

This conversion can be defined as a change of relative coordinates, from real spatial and angular coordinates at the takeoff $(x_0, \theta_0,)$ to virtual coordinates $(x, \theta)$ [36]. Given the overall lateral magnification of the objective lens M and defined the immersion factor as $k = E/E_0$, one has the following relations:

$$M = \frac{2}{3} M_M, \quad M_A = \left( k^{\frac{1}{2}} M \right)^{-1}$$

$$\frac{x}{x_0} = M, \quad \frac{\theta}{\theta_0} = M_A, \quad \frac{q_0}{q} = M,$$

whereas $q_0 = \theta_0/\lambda_0$ and $q = \theta/\lambda$ are the spatial frequencies parallel to the surface in the real and virtual plane respectively, and $M_A$ is the angular magnification. In the case of low energy electrons, i.e. if $E_0$ is just a few eV, the application of a potential of few tens of kV guarantees a large immersion factor.

In the simplest configuration, the objective lens displays the magnified image to the electron detector, e.g., a fosforescent screen. In this case, only a limited magnification can be reached: the overall lateral magnification of the modern objective lens ranges typically between 15 and 40. To increase the performances of the microscope, a series of magnetic and electrostatic lenses can be added, to constitute the so-called *imaging column*. The focal length of the lenses can be controlled by changing the current of magnetic lenses, or the potential of electrostatic lenses. Electrons travel through the imaging column with kinetic energy equal to $E = E_0 + eV$. Since usually electrons can escape from a surface with different $E_0$ simultaneously, their trajectory in the imaging column can differ considerably from the optimum one. To guarantee that electrons with a selected initial kinetic energy pass through the lenses along the optimal trajectory, a tunable bias must be subtracted from the potential $V$ between sample and aperture. The bias value, also called *start voltage* ($V_0$), imposes the optimal trajectory and speed to only the electrons with initial kinetic energy $E_0 = eV_0$. For them, the final kinetic energy will be $E = eV$. Electrons with different kinetic energy still pass through the objective lens, but can be easily filtered on a later stage due to their different trajectories. Moreover, the fixed trajectory and speed of the electrons allow to set the lenses of the imaging column once and for all. This not only improves the usability of the instrument, but permits to place apertures and slits to mould the image in a convenient plane.

The advantage of having an imaging column is twofold. Besides the improved magnification, it gives access to the angular distribution of the emitted electrons. In optics, for objects a finite distance away, rays that leave the object with a given angle (**Fig. 2b**, orange trajectory) cross at a precise point in the backfocal plane of the objective lens. There, the image maps the distribution of electrons as a function of their emission angle, i.e. in the reciprocal space. In the case of electrons backscattered or photoemitted from a crystalline surface, this plane contains the diffraction pattern. The imaging column can then be set to display a magnification of the backfocal plane on the screen instead of the image plane.

At the end of the imaging column, it is possible to place an *energy filter*, that exclude electrons with different kinetic energy. By knowing the displacement as function of energy, one can filter out the electrons by placing a slit of given size. Typically, the energy analyzers is optically neutral, i.e. the entrance plane is displayed in the dispersive plane at the exit with unit magnification. Finally, another series of lenses project the desired plane onto the detector.

In conclusion, three different operation modes can be defined:

- In the *Microscopy* mode (**Fig. 3a**) the image plane is displayed on the screen. In the imaging column the backfocal plane is reproduced and an aperture (called *contrast aperture*) can be inserted to limit the acceptance angle of the electrons. If an energy analyzer is installed, the imaging column is set to display the reciprocal plane at its entrance. The energy slit is placed at the dispersive plane and lets pass only electrons with a selected kinetic energy. The projector displays back the image plane onto the detector. Since the apertures are inserted on diffraction planes, the real image is still fully available.
- The *Diffraction* mode (**Fig. 3b**) displays the distribution of electrons in the reciprocal space. In this case no contrast aperture has to be placed. Nonetheless, an aperture in the image plane of the objective lens, called *field limiting aperture*, can filter the electrons in the real space: the diffraction image is then made only from electrons emitted in a particular area. The imaging column transfers the image plane to the entrance of the energy filter: the slit at the dispersive plane does not influence the reciprocal image. The projector converts the image plane at the exit of the analyzer to the diffraction plane.
- In the *Spectroscopy* mode (**Fig. 3c**) the projector magnifies the dispersive plane at the end of the energy filter. In this case both the contrast aperture and the field limiting aperture can be inserted, to limit conveniently the acceptance angle and the probed area. At the entrance of the energy analyzer the reciprocal plane is usually displayed. The dispersive plane looks then as a line with modulated intensity. The intensity line profile over the spreaded electron beam reveals the energy distribution of the electrons.

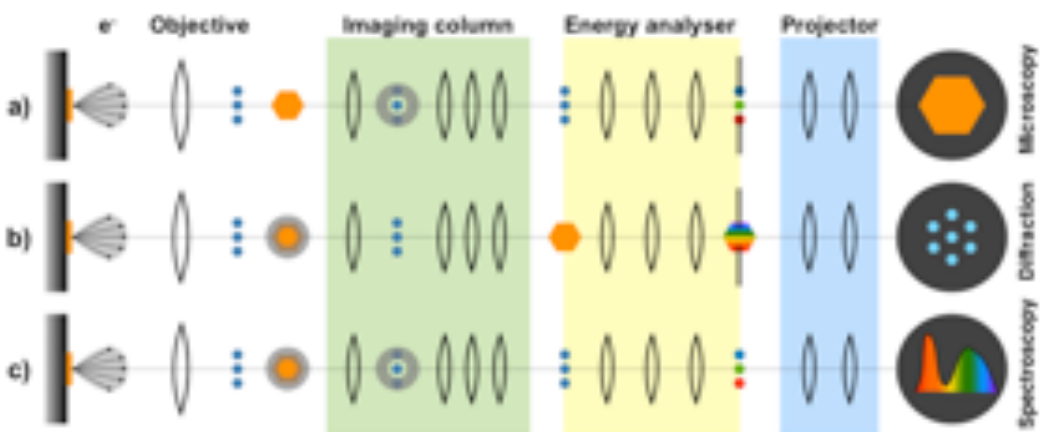

Figure 3: Scheme of PEEM optics with energy filtering. Three main operational modes are presented: (a) Microscopy mode, (b) Diffraction mode, (c) Spectroscopy mode. The hexagon represents the image plane, while the blue dots the diffraction plane. Rainbow colors symbolize the dispersive plane.

The easy switch between the three operational modes is at the origin of the versatility of cathode lens microscopy. In addition, the insertion of apertures and slits in the convenient planes allows to combine microscopic, spectroscopic and diffraction information in a single experiment: the active filtering in real space, reciprocal space and kinetic energy can be simultaneously activated to obtain a unique characterization of the probed surface.

## 2.2 Instrumentation

The simplest setup, with objective lens, imaging column and electron detector, is already capable of performing microscopy and diffraction measurements on surfaces. Over the years, cathode lens microscopes have become more sophisticated, with the addition of multiple optical elements. This section will show the most common components available today, with brief discussion on the working principles and the experimental advantages introduced.

### 2.2.1 Beam separator

The beam separator is a magnetic element that imposes different trajectories to electron beams with opposite direction by using the Lorentz force. While in standard PEEM it has no practical use, it is an essential element in LEEM and in aberration-corrected instruments equipped with an electrostatic mirror. The decoupling between incident and reflected electrons allows a separate treatment of the two beams, i.e. in LEEM a full-field detection of the illuminated area, with no shadows casted by the electron gun (like for standard LEED optics). The beam splitter is usually placed after the objective lens, so apertures and slits can be introduced along the incident or the reflected path separately.

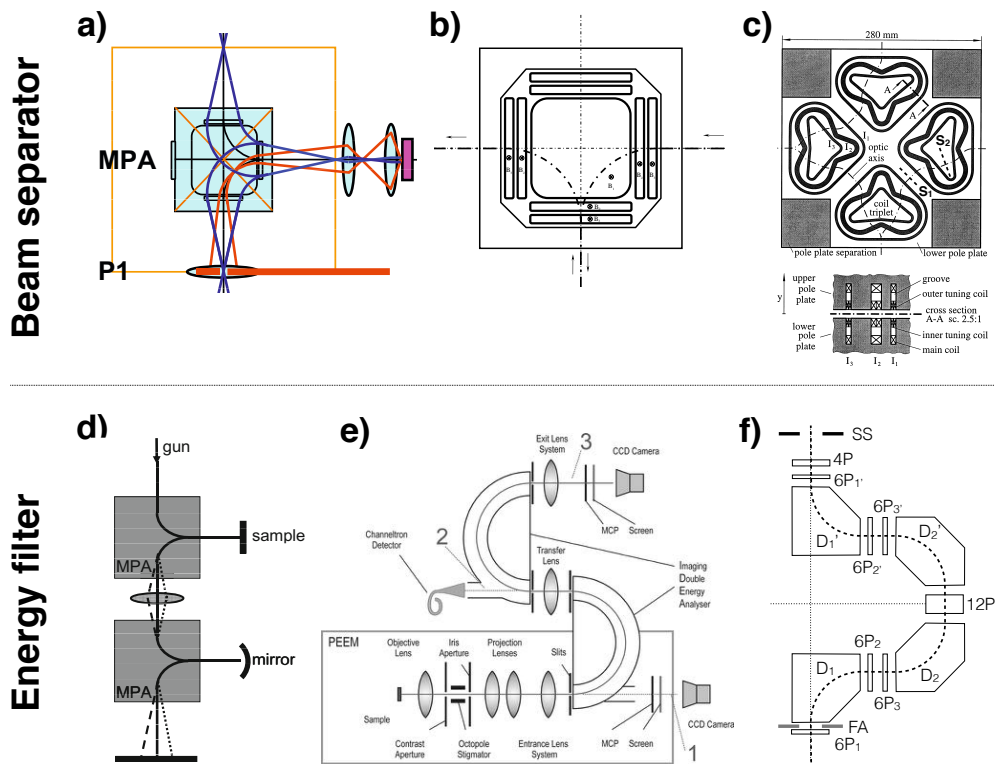

Figure 4: Scheme of beam separators (top) and energy filters (bottom) used in cathode lens microscopy. (a) Diagram of the Magnetic Prims Array used in Tromp's IBM LEEM-II system. The orange boxes around the MPA indicate the position of diffraction planes. Intermediate image planes are located on the MPA diagonals. Reproduced from Ref. [26] with permission, copyright 2010 Elsevier. (b) Layout of Mankos' Magnetic Prism Array with four pairs of rectangular sectors. Reproduced from Ref. [37] with permission, copyright 2007 Elsevier. (c) Midsection view of the twofold-symmetric beam separator that equips SMART. Vertical cross section along A-A is represented below. Symmetry plane S1 and S2 are highlighted. Reproduced from Ref. [25] with permission, copyright 1997 Elsevier. (d) Diagrams of energy filtering with a double MPA. The induced dispersion is depicted with dashed lines. Reproduced from Ref. [26] with permission, copyright 2010 Elsevier. (e) Schematic layout of the NanoESCA instrument equipped with Imaging Double Energy Analyzer. The three path of electrons are indicated: (1) PEEM mode, (2) selected-area spectroscopy and (3) energy-filtered ESCA imaging. Reproduced from Ref. [38] with permission, copyright 2005 IOP Science. (f) Schematic layout of the Omega filter equipping SMART. Dipoles (D), quadrupoles (4P), hexapoles (6P) and dodecapoles (12P) are highlighted. The introduction of a Field Aperture (FA) and a Selection Slit (SS) enables the filtering. Reproduced from Ref. [25] with permission, copyright 1997 Elsevier.

The first separator installed in the original LEEM was a simple Archard-Mulvey type with 10° deflection [39]. Very soon it became clear that the magnetic field strongly impacts the image quality [16], therefore more sophisticated and multipolar magnetic prism arrays were developed. Over the years, two geometries emerged as standard: 120° deflection (Bauer/Elmitec) and 90° deflection (Tromp/SPECS, SMART, PEEM3 and others). For the first case, unfortunately no detailed information has been published. In the other case, several solutions have been employed. In the square magnetic prism array (MPA) used by Tromp (**Fig. 4a**) a large central squared and four rectangular magnetic field segments provide the deflection and a stigmatic refocus of the electron beam [26, 40], so that the images produced at the entrance and exit plane are equivalent. The lenses of the microscope are set to display the focused incoming electron beam and the outgoing diffraction plane at the entrance planes. With this geometry, the image plane is created on the diagonal plane of MPA. The contrast aperture is then displaced in the exit plane of MPA, together with the first projective lens. A similar design, but with more elements, is used in a LEEM with dedicated optics for high-throughput performances (**Fig. 4b**) [37]. In this case, four pairs of smaller rectangular sectors, in which the magnetic field is about three times stronger, surround the central squared magnetic field. The advantage to have a pair of independent coils per side is that the same 90° deflection can be achieved with different pairs of flux density values, favoring a more precise alignment of the device and allowing a larger field of view without significant distortions.

It is worth noticing that the Lorentz force imposes different deflection to electrons with different kinetic energy, i.e. the MPA displays a chromatic dispersion on the exit plane. As will be discussed in Sect. 2.2.2, the beam deflector can be used as an energy filter. Moreover, the four sectors can be independently set to deploy different magnetic field strengths, so to deflect electron beams when the incoming and outgoing electrons have different kinetic energy. This is the case of Secondary Electron Microscopy (SEM) and Auger Electron Microscopy (AEM), which use slow secondary electrons or Auger electrons as information carriers. In both cases, the kinetic energy of the emitted electrons is lower than the one of the electrons used for illumination: the beam separator must then be set to deflect properly both beams along the optimum trajectory.

The performances of beam separators became more crucial with the advent of aberration-corrected instruments. In LEEM/PEEM systems, the compensation of aberrations is made with the introduction of a particular electrostatic mirror along the electron path (see Sect. 2.2.3). Its application requires a second deflection that separates entering and reflected beams. The beam separator used for this purpose needs not only to be stigmatic and distortion-free, but also with no chromatic dispersion. To do so, two solutions have been employed. The first is to use two identical deflectors, one in front of the objective lens and one for the electrostatic mirror, connected by transfer optics. In this case, most of aberrations and the chromatic dispersion are cancelled out by symmetry. Rose and Preikszas proposed a second solution with the development of a highly symmetric four-quadrant beam separator (**Fig. 4c**) [41], used successfully by SMART and PEEM3 instruments [25, 27, 42]. Each quadrant contains a coil triplet that produces two regions with opposite magnetic field. Electrons pass through the field of two coils for every 90° deflection. The shape of the triplet is such that the fields are axially symmetric and point symmetric about the diagonal planes of the beam separator ($S_1$) and the bisector plane of the coil triplet ($S_2$). This double symmetry ensures automatic compensation of deviations and dispersion up to the second order. Since the Rose deflector has four quadrants, it can be used for both deflections required by the objective lens and the mirror.

## 2.2.2 Energy Analyzers

As seen in Section 2.1, the energy filtering of electrons activates a third, scientifically very important operating mode of cathode lens microscopy. The most logical way to filter electrons with different kinetic energy is to exploit their different trajectory along the optical path and to cut them out with an aperture. The energy window ΔE around the pass energy E required to perform active spectroscopy of photoemitted electrons is typically below 200 meV; since E is some tens of keV, the order of magnitude of the resolving power E/ΔE must be considerable, about $10^5$. The simplest device that acts as energy filter is the beam separator. The brightest example of this kind is the aberration-corrected LEEM developed by Tromp, equipped with two identical deflectors (**Fig. 4d**) [26]. In this setup, the MPAs are connected by a transfer lens, which resends the dispersed image at the exit of MPA1 to MPA2. This double pass guarantees an achromat image on the electrostatic mirror. After a second pass through MPA2, the newly formed dispersive plane is used to filter electrons on a narrow energy window with a slit. The low dispersion of the beam separator, 6 μm/eV, allows a proved energy resolution of 250 meV.

To achieve better resolution performances, a dedicated energy analyzer with larger dispersion is necessary. The fact to have a separate device to filter electrons has some further advantages. In fact, despite the more complicated setup, it allows full control over the energy window and the pass energy. Moreover, it enables active filtering in the *diffraction* mode if the image plane is projected at its entrance, as mentioned in Section 2.1. The first spectroscopic instrument, the SPELEEM [21, 43], was equipped with an electrostatic hemispherical deflector analyzer (HDA), a common solution for filtering in photoelectron spectroscopy. In the analyzer electrons travel through the space between two concentric hemispheres held at different potentials. The electrostatic field disperses the electrons depending on their kinetic energy around an optimal trajectory, given by particles with a well defined *pass energy*. The resolving power $E/\Delta E$ of an HDA is typically $10^3$-$10^4$; therefore for cathode lens microscopy it is necessary to slow down electrons with a dedicated retarding lens from E (few tens of keV) to a pass energy of 1 keV or less. Such deceleration is a critical parameter, since the resulting immersion factor expands the angular spreading of the electron beam and degrades the lateral and the energy resolution via spherical aberrations. In the first version of SPELEEM the pass energy in HDA was 1800 eV and the demonstrated energy resolution was 0.5 eV. The optimized commercial version by Elmitec lowered the pass energy to 900 eV to obtain a reported energy resolution of 110 meV in *spectroscopic* mode and about 150-200 meV in the other modes [44]. The passage through an HDA induces second-order aberrations at the exit plane. Their correction can be achieved with the introduction of a second twin HDA (**Fig. 4e**). This configuration, called Imaging Double Energy Analyzer (IDEA), equips the NanoESCA PEEM [38, 45]. The symmetry of path forces the electron trajectories to coincide after the double passage, thus generating an achromat image at the exit plane. The energy filtering in *imaging* and *diffraction* mode is obtained with the introduction of a slit in the dispersive plane placed between the HDAs. This system is capable of a demonstrated energy resolution of 12 meV with pass energy 15 eV. Higher pass energies, more suited for core-level spectroscopy and imaging, degrades the resolution to 50-100 meV. Remarkably, NanoESCA can also work as a single-pass photoelectron energy analyzer and as a classic PEEM with no energy filtering.

Another filtering solution with large dispersion factors is the so-called "Omega filter" that was originally developed for TEM [46] and that now equips the SMART instrument (**Fig. 4f**) [25, 42]. It is made by four magnetic 90° deflectors, arranged in a way that the resulting optic axis resembles the Greek capital letter Omega. The symmetry of the path and the placement of a quadrupole, six hexapoles and a dodecapole on convenient planes allow correction of all second-rank aberrations. The pass energy of this instrument is 15 keV, i.e. no retarding field is required, and the designed resolving power is 150000. The calculated dispersion at the exit plane, 35 μm/eV, is large enough to display a window of ∼ 10 eV in *spectroscopy* mode with a demonstrated energy resolution better than 180 meV [47].

### 2.2.3 Aberration correctors

In optics, aberration is the deviation from the nominal image raised by defects of the optical system. Such deviation can depend on geometrical factors (rays with different initial trajectories can be refocused on different planes - *spherical aberration*) or physical factors (the refraction index of the lens changes with the wavelength of the ray - *chromatic aberration*). In light optics, an easy way to correct aberrations is to combine convex and concave lenses conveniently, since the two types produce aberrations of opposite sign and the overall effect can cancel out. In electron optics, this circumstance is prohibited by the Scherzer's theorem [48]:

> *The chromatic and spherical aberrations of an electron microscope with round lenses, real images, static fields, no space charge and a potential and its derivative without discontinuities, are always positive.*

The resolution of a cathode lens microscope is then dominated by chromatic and spherical aberrations, mainly resulting from the objective lens. In the ideal case of an aberration-free system, the image of a point-like source is again a point. The blur induced by aberrations transforms the point image into a disk with a width $d_0$. Since effects like coma and field distortion are negligible in LEEM/PEEM system, $d_0$ can be expressed as a Gaussian convolution of the contributions given by chromatic aberration, spherical aberration and the diffraction limit [49, 50]. Given the acceptance angle $\alpha$, the energy width ΔE and the start energy $E_0$, one has that

$$d_0 = \sqrt{d_d^2 + d_s^2 + d_c^2} \ ,$$

with

$$d_d = \frac{0.61\sqrt{\frac{1.5}{E_0}}}{\sin \alpha}$$

$$d_s = C_0 \sin \alpha + C_s \sin^3 \alpha + C_{ss} \sin^5 \alpha$$

$$d_c = C_c \frac{\Delta E}{E_0} \sin \alpha + C_{cc}\left(\frac{\Delta E}{E_0}\right)^2 \sin \alpha + C_{sc} \frac{\Delta E}{E_0} \sin^3 \alpha$$

Here, $d_d$ is the radius of the confusion spot due to diffraction at the smallest aperture. The other components $d_s$ and $d_c$ are the radii of the confusion disc due to spherical and chromatic aberration, expressed to the lowest orders of a Taylor series. In standard experimental conditions $d_d$ is less than 1 nm, thus the resolution is mainly determined by spherical and chromatic aberrations. The effect of the first one is predominant at higher kinetic energy, while the second is more significant at lower kinetic energy.

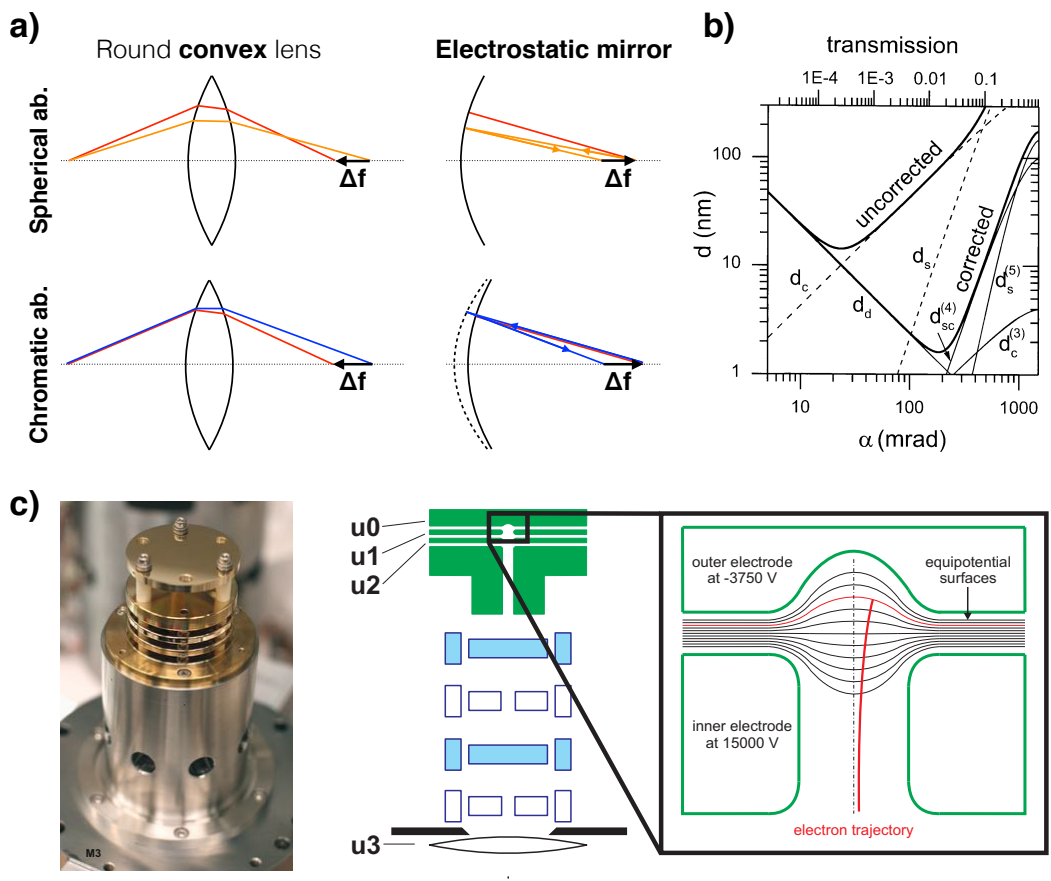

Figure 5: (a) Schematics of the effect of spherical and chromatic aberration on a round convex lens and on an electrostatic mirror. (b) Resolution limit as a function of the acceptance angle α for uncorrected and corrected SMART in the case $E_0$ = 10 eV and ΔE = 2 eV. The dominating aberration components are added: dashed line for the uncorrected and thin solid lines for the corrected case. Reproduced from Ref. [49] with permission, copyright 2002 World Scientific. (c) Photograph of the electron tetrode mirror assembly used in IBM LEEM-II and scheme of the tetrode mirror that equips SMART. The equipotential surfaces in the latter mirror stage are highlighted. Reproduced from Refs. [26] and [25] with permission, copyright 2010 and 1997 Elsevier.

The constraints of Scherzer's theorem can be circumvented in many ways: some solutions were already known in the early years of theoretical electron optics, but were not implemented before the 1970's [51, 52]. Historically, the field of aberration correction was pioneered by TEM community [53, 54], while its development in cathode lens microscopy started much later. Among the multiple methods already tested in TEM, the most convincing one for cathode lens microscopy is the use of an electrostatic mirror [41, 55–57]. The principle how a mirror can compensate spherical and chromatic aberrations is shown in **Fig. 5a**. In the first case, an electrostatic mirror and a round convex lens with the same radius produce focal displacements $\Delta F$ of equal magnitude and opposite of sign. In the second case, the electrostatic mirror deflects more the trajectory electrons with higher kinetic energy, compensating the focal displacement $\Delta F$ induced in the round convex lens. The electrostatic tetrode mirror can compensate the aberration effect by reducing simultaneously the lower order coefficients $C_s$ and $C_c$ to zero. The effect on resolution and transmission calculated for the SMART instrument is presented in **Fig. 5b**. The cancellation of low-order aberration coefficients can improve the resolution by an order of magnitude. Moreover, since the acceptance angle can now be broaden without loss of resolution, the transmission of the microscope results enhanced as well, with sensible reduction of the acquisition time.

The mirror is currently employed in several aberration-corrected systems. The first in chronological order is a pure PEEM microscope equipped with a hyperbolic mirror with only two electrodes, capable of compensating simultaneously spherical and chromatic aberrations only for one magnification and one energy [58, 59]. The maximum flexibility is obtained with a tetrode mirror (**Fig. 5c**), currently employed in several instruments (SMART, PEEM3, Tromp/SPECS, Elmitec) [25–27]. While the first electrode is at ground potential, the potential of the three others can be varied. They modify the shape of the equipotential surfaces that act as a mirror for the incoming electrons, thus determining the focal length, the chromatic aberration and the spherical aberration. The mirror is then set conveniently with the operation mode and kinetic energy to cancel out the primary aberration coefficients induced by the lens system. The $C_s$ and $C_c$ coefficients of the mirror can be calculated in a reasonable amount of time for a given $E_0$, so that standard values of the electrodes can be easily set. Furthermore, the aberration coefficients of the system can be directly evaluated with a series of routine measurements, thus enabling fine correction. PEEM3 reported a lateral resolution of 5.4 nm for PEEM images of biological samples with a 233-nm laser as photon source [59]. The first aberration-corrected LEEM image of SMART visualized nanometer surface structures such as the herringbone reconstruction of the Au(111) surface with a lateral resolution of 2.6 nm [60]. Both the Tromp/SPECS and Elmitec systems report now an ultimate lateral resolution below 2 nm in LEEM [26]. Better results are still theoretically possible, but are very difficult to achieve for long periods of time due to the intrinsic instability of the corrected state, which constrains the lifetime of the corrected state to just a few minutes [61]. Nonetheless, it should be remarked that the gain in transmission guaranteed by the aberration corrector is of great help for measurements, e.g., with X-ray photoemitted electrons, where the exposure time and the weakness of the photon source can be a crucial issue for the success of the experiment [62].

### 2.2.4 Electron and Photon Sources

The electron source in LEEM is an electron gun capable of emitting electrons with high brilliance and narrow energy distribution. During the years, several kinds of emitters were used. The most frequent electron source is a $LaB_6$ or $CeB_6$ crystal with conical shape showing the (100) surface on the flat tip. Once heated by a filament, electrons leave the crystal via thermoionic emission from the tip, due to the low work function of the (100) surface. A Wehnelt aperture placed in front of the tip with negative potential suppresses the emission from other faces of the crystal [63]. These thermoionic emitters have a long lifetime and can draw a very high current with an energy width larger than 0.7 eV [64]. The use of thermoionic emitters is not recommended for high resolution microscopes: cold field emitters and Schottky emitters can generate electron beams with a narrower energy spread (0.3 eV), thus reducing the influence of chromatic aberrations, at the cost of lower brilliance and shorter lifetime [65–67]. More rare, but with very interesting applications, are the spin-polarized electron sources. The interaction between spin-polarized electrons and the specimen in LEEM systems provides unique information on magnetic phenomena with lateral resolution (see Sect. 3.1.4). The most common one uses the photoemission of electrons from III-V semiconductors with circularly polarized light [68–71]. Under particular conditions of strain, the photocathode can generate an electron beam with polarization ~ 0.9, while the selection of the light polarization switches easily the polarization vector of the beam.

The photon sources employed in PEEM range on the wavelength of photons and the time structure of the light pulses, enabling a wide variety of surface science experiments. The easiest way to produce photoemission is with continuous UV illumination by discharging lamps. The first PEEM systems used a Hg Short-Arc lamp emitting UV light at 4.9 eV (254 nm). The low energy of the photons restricted its application to samples with low work function and low electron affinity. Higher photon energy can be provided by a He gas discharge lamp: the HeI emission mode (21.6 eV) is still used for energy-filtered angle-resolved photoemission diffraction measurements of valence band electrons.

Laser sources were used to stimulate photoemission in PEEM since the 1970s [72]. Pulsed lasers can provide very short flashes of light along a wide wavelength spectrum and are currently used for time-resolved studies in pump-probe mode and for multi-photon photoemission. Several systems are suited for LEEM, such as Nd:YAG and Ti:Sapphire, and are often used with higher harmonic photon generation [73–75]. The high throughput of lasers concentrated in a single pulse is the major limitation to their employment in

PEEM measurement. The dense bunch of photoemitted electrons experience a reciprocal Coulomb repulsion during their travel to the detector, causing a general degradation of the carried information (the so-called "space charge" effect – see Sect. 4.0.2). This phenomenon can be mitigated with a high pulse repetition rate and with a low energy per pulse, so that PEEM experiments with a dynamic timescale of some tens of fs and good lateral resolution are now possible [76].

The most successful photoelectron source for PEEM is synchrotron radiation. Since the first installation of SPELEEM in Elettra, Italy, it was clear that the tunable, intense light provided by insertion devices in third-generation synchrotron is the most versatile and powerful complement to cathode lens systems [21]. For example, the SPELEEM beamline [44, 77] is equipped with two Sasaki Apple II undulators that provide elliptically polarized light (circular left and right, linear horizontal and vertical) in a spectral range between 40 and 1000 eV. The light is monochromatized by two Variable Line Space plane gratings with a resolving power $E/\Delta E$ = 4000 at 400 eV. The beamline flux exceeds $10^{13}$ photons at 150 eV and is above $10^{12}$ photons in an energy range between 50 and 600 eV. These numbers permit to carry a great variety of experiments with valence band and core-level photoemitted electrons and with state-of-the-art lateral and energy resolution. The only technical limitation arises by the pulsed structure of the synchrotron radiation: space-charge effects limit again the lateral resolution of XPEEM images with core-level electrons to about 20 nm [62, 78]. This restriction could be overcome not only with photon intensity reduction, but also with an intelligent placement of apertures to cut away electrons not used for imaging. Nowadays, about 20 synchrotron endstations are equipped with a PEEM, with energies ranging from near-UV to hard X-rays (HAXPEEM), giving the biggest contribution to the growth of cathode lens microscopy user community.

## 2.3 Performances

At this point it is useful to summarize the performances of the various imaging techniques of cathode lens microscopy, highlighting the advantages and underlining the limitations. The list incorporates some of the state-of-the-art results, as well as routine performances achievable with good quality samples. The purpose of this section is to help the non-expert reader to choose the right technique that fits their needs, with no claim to be exhaustive. It must be remarked that the state-of-the-art results are obtained in very controlled and stable environmental conditions, with flat and conductive samples and cannot be achieved in every measurement of that kind. A charging and non-atomically flat sample can degrade the performances easily by an order of magnitude. In this sense, the routine values are more significant, as they give a more realistic expectation for a generic experiment.

| | | | |
|---|---|---|---|
| Lateral resolution | LEEM | | |
| | non-aberration corrected | Routine: 20 nm<br>Best: 4.1 nm (LEEM)<br>(SPLEEM) | [26] |
| | aberration corrected | Routine: 5-10 nm<br>Best: 2.6 nm<br>2.0 nm<br>2.0 nm | [26, 30, 60] |
| | PEEM | | |
| | non-aberration corrected | Routine: 40-100 nm<br>Best: 7.0 nm | [33] |
| | aberration corrected | Routine: 40 nm<br>Best 5.4 nm (UVPEEM)<br>18 nm (XPEEM)<br>2.6 nm (laser) | [59, 62, 79] |
| Energy resolution | PEEM | | |
| | with MPA energy filter | Routine: 1-2 eV<br>Best: 0.25 eV (spectroscopy)<br>1.7 eV (imaging) | [31] |
| | with HDA energy filter | Routine: 0.7 eV<br>Best: 0.11 eV (spectroscopy)<br>0.2 eV (imaging) | [30, 44] |
| | with double HDA | Routine: 0.1 eV<br>Best: 0.01-0.05 eV | [80] |
| Angular resolution | with HDA | Best: 0.047 $\text{Å}^{-1}$ | [44] |
| | with double HDA | Best: 0.005 $\text{Å}^{-1}$ | [80] |
| Time resolution | laser | Routine: tens of fs (lateral resolution 20-50 nm)<br>Best: 200 as (lateral resolution 200 nm) | [81, 82] |
| | synchrotron | Few ps (single bunch width) | |

**Table 1: list of performances for the various operating modes of cathode lens microscopy**

## 3. Low Energy Electron Microscopy

LEEM uses backscattered electrons as information carriers. Unlike scanning microscopy, electrons are collected simultaneously from an illuminated area of several tens of μm. The image formed by the magnification lenses can then be acquired even in video-rate (down to 1 ms/frame), depending on the detector quality and the signal intensity. The image contrast depends on how electrons interact with the surface: the higher or lower reflectivity can depend on several factors, e.g morphology, crystallinity and quantum effects. It is therefore important to understand how electrons interact with the surface and how the image is formed. Elastic and inelastic scattering of electrons on solids is a well-studied subject in condensed matter physics. Here will be summoned only the most important concepts, leaving a more complete description to other textbooks [83].

### 3.0.1 Basic Image Contrast

The simplest conceptual case from where to start is the "single scattering" frame, i.e an electron scattered only once by a surface atom. Here, the scattering amplitude is given by the atomic scattering factor $f_n(s)$, where $s = k_{out} - k_{in}$ is the momentum transfer between incident and diffracted plane wave with wave vectors $k_{in}$ and $k_{out}$, respectively. Considering now a monoenergetic electron beam, represented by a plane wave with amplitude

$$\psi_{in} = \psi_0 e^{ik_{in}\cdot R} \quad ,$$

the amplitude of a diffracted beam is represented by

$$\psi_{out} = \psi_0 \left[\sum_n \alpha f_n(s) e^{is\cdot R_n}\right] e^{ik_{out}\cdot R}$$

Here $f_n(s)$ is the atomic scattering factor for the $n$th atom at position $R_n$. For an elastic scattering, the kinetic energy E₀ must be preserved, i.e.

$$E_0 = \frac{\hbar^2}{2m}|k_{in}|^2 = \frac{\hbar^2}{2m}|k_{out}|^2$$

The contribution of atomic scattering factor and the diffraction effects generate image contrast in LEEM. In fact, areas with different composition, stoichiometry and crystal structure will have a different electron reflectivity and will appear in LEEM as brighter and darker areas.

For a complete description of electron reflection, inelastic effects and multiple scattering must be taken into account. When traveling inside solids, electrons have a certain probability to experience an inelastic event. Therefore, their probability to be reflected with no losses depends on how deep the scattering center is placed into the bulk. To model this behavior one can introduce a mean free path expressed by an imaginary component of the electron-surface interaction potential, such that the scattering amplitude decays exponentially in the direction of wave propagation. In general, the electron mean free path is energy dependent and relatively independent of the material, so that its value follows a "universal curve". Such universal curve has a V-shape, i.e. presents a minimum for energies around 30-100 eV: in this range the mean free path is so small (few Å) that elastic electrons come from only the top-most atomic layers. LEEM and PEEM performed in this range are then surface sensitive. Electrons with higher kinetic energy can probe the sample more in depth, while at very low energy (~ 10 eV) the inelastic mean free path can show large deviations accordingly to the density of states of the material: at a few eV, organic thin films with very low density of states can show a mean free path of 10 nm, while transition metals with dense *d* or *f* bands above the Fermi energy can damp scattered electrons already at a depth of 0.5 nm. The surface sensitivity can be used in LEEM to achieve image contrast even between samples with the same stoichiometry and different surface reconstruction.

At very low kinetic energy, another quantum phenomenon can affect the elastic backscattering of electrons from thin films. When electron wavelength and penetration depth are comparable to film thickness, i.e. at very low kinetic energy, the confinement imposed by the vacuum boundary and the film-substrate interface induces a one-dimensional quantum well condition [84]. This so-called quantum size effect (QSE) rises from the interference between electron waves reflected at the surface and at the film-substrate interface. In first approximation, for a film of thickness *d*, the phase shift induced by the different path length is

$$\phi = \left(\frac{2d}{\hbar}\right)\sqrt{2m(E_0 + V_i)} \quad [1],$$

where $V_i$ is the inner potential of the thin film. The electron reflectivity is then subject to periodic oscillations as a function of electron kinetic energy and film thickness. QSE is extensively used to measure directly the thickness of thin films: significant examples will be given in Sect. 3.1.1.

### 3.0.2 Image Formation

The reflection of plane waves is influenced also by the morphology of the surface: atomic steps, kinks, domain boundaries and defects create interference and modulate the electron reflectivity. To better understand how to interpret the features in a LEEM image, it is necessary to address the theory of image formation. Over the last two decades, several approaches were used to calculate the image formation in LEEM. The first is from Chung and Altman [22, 85], who developed a wave-optical model to describe the step contrast in ideal and real conditions, i.e. taking into account instrumental broadening and beam coherence. Later the model was improved by a Fourier Optics formalism [36], which elucidates the image formation for objects with different scattering amplitude and phase and incorporates aberration effects of the objective lens, diffraction cut-off by a contrast aperture, lens defocus, energy spread of the electron beam and instabilities in lens current and voltage. In parallel, Jesson and coworkers [86, 87] proposed an alternative approach, based on the definition of a Contrast Transfer Function (CTF), into which flow all the effects of the imaging system on the transfer from real object to image. Schramm et al. [88] integrated this method with fifth-order aberrations, making it suitable for aberration-corrected instruments. The CTF formalism is attractive for its low computational cost and its universal treatment of arbitrary phase, amplitude or mixed amplitude-phase objects. In the following, a brief excursus on the CTF formalism is depicted.

Consider having an object illuminated by a monoenergetic plane wave, as in Sect. 3.0.1. It has been shown that the reflection causes variation of wave amplitude and phase. In general, the reflected wave is given by

$$\psi_{out} = \psi_{in} * \psi_{obj}$$

with

$$\psi_{obj}(\bm{R}) = \sigma(\bm{R})e^{i(\bm{k}_{out}-\bm{k}_{in})\cdot\bm{R}}e^{i\phi(\bm{R})}$$

Here $\sigma(\bm{R})$ is the amplitude modification factor, while the phase modification factor $\phi(\bm{R})$ incorporates the phase difference between outgoing and incoming waves induced by the surface morphology. Supposing that the surface is the plane $xy$ at $z = 0$, its morphology can be modeled by a surface height function $h(\bm{R_0})$ expressed in unit of step height $a_0$. $\bm{R_0}$ is then a two-dimensional position vector spanning over the surface plane. In case of normal incidence, only the vertical component of the wave vector matters, i.e. $\bm{k} = |\bm{k}|\hat{z}$. Given $|\bm{k}| = 2\pi/\lambda_0$, the phase shift defining the surface is given by

$$\phi(\bm{R_0}) = 2\frac{2\pi}{\lambda_0}a_0\, h(\bm{R_0})$$

The phase object function includes the effects of surface morphology on the reflected wave, assuming no significant modification of the accelerating electric field.

The reflected wave is then modified by the cathode immersion lens. First of all, the acceleration from kinetic energy $E_0$ to $E$ imposes the change of coordinates as in Section 2.1, from takeoff coordinates in real ($\bm{R_0}$) and reciprocal space ($\bm{q_0}$) to respective virtual coordinates ($\bm{R}$ and $\bm{q}$). The transfer from virtual object to magnified image can be described in real or reciprocal space. In real space, the response of the system is described by the Point Spread Function (PSF) $T(\bm{R})$, which models the blurring of an ideal point object. The final image $\psi(\bm{R})$ is then the convolution of the outgoing wave $\psi_{out}$ and the PSF,

$$\psi(\bm{R}) = (\psi_{out} \otimes T)(\bm{R})$$

Using the fact that convolution in real space corresponds to multiplication in the Fourier space,

$$\psi(\bm{R}) = \mathcal{F}^{-1}\big[\mathcal{F}[\psi_{out}(\bm{R})] * \mathcal{F}[T(\bm{R})]\big]$$

$T(\bm{q}) = \mathcal{F}[T(\bm{R})]$ is the Contrast Transfer Function (CTF) of the LEEM imaging system and it is modeled as the product of all relevant contributions imposed by the optical system, i.e. the frequency cutoff imposed by the contrast aperture, chromatic and spherical aberration, defocus and instrument-related instabilities. One can model $T(\bm{q})$ as follows:

$$T(\boldsymbol{q}) = M(\boldsymbol{q})\, W(\boldsymbol{q}, \Delta f)\, \mathcal{E}(\boldsymbol{q}, \Delta E)$$

Here, $M(\boldsymbol{q})$ incorporates the effect of a round contrast aperture placed in the backfocal plane, where the reciprocal space is displayed. Its effect is to filter high spatial frequencies

$$M(\boldsymbol{q}) = \begin{cases} 1 & if\ |\boldsymbol{q}| < q_{max} \\ 0 & if\ |\boldsymbol{q}| \geq q_{max} \end{cases}$$

$q_{max}$ corresponds to the maximum spatial frequency imposed by the aperture size and it is equal to $\alpha_{max}/\lambda$, being $\alpha_{max}$ the maximum angle accepted. It should be noticed that the contrast aperture acts downline of the acceleration stage, therefore the electron wavelength is calculated from the final kinetic energy $E$.

The wave aberration contribution $W(\boldsymbol{q})$ refers to deviations of the wave path from the ideal one, induced by defocus $\Delta f$ and by spherical aberrations, which may be expressed by Taylor series coefficients as in Sect. 2.2.3

$$W(\boldsymbol{q}, \Delta f) = exp\left(\frac{i\pi}{2}\left(C_s \lambda^3 q^4 + \frac{C_{ss}}{3}\lambda^5 q^6 - 2\Delta f \lambda q^2\right)\right)$$

The defocus $\Delta f$ takes into account also the unintentional focus oscillations caused by voltage and current fluctuations in lenses and high voltage supplies.

Finally, the chromatic aberration damping envelope $\mathcal{E}(\boldsymbol{q}, \Delta E)$ comes from an integration over the weighted contribution of the different energies within the Gaussian energy distribution with FWHM $\Delta E$. Limiting the expression of chromatic aberration to the first-order coefficient, one has that

$$\mathcal{E}(\boldsymbol{q}, \Delta E) = exp\left(-\frac{(\pi C_c \lambda q^2)^2}{16\, ln2}\left(\frac{\Delta E}{E}\right)^2\right)$$

The final LEEM image is an intensity distribution of the reflected wave modified by the lens system, so it can be calculated as

$$I(\boldsymbol{R'}) = \frac{1}{M^2}|\psi(\boldsymbol{R})|^2\ ,$$

$\boldsymbol{R'}$ being the two-dimensional coordinates at the detector plane.

The CTF approach is currently used to produce simulation of LEEM images for a surface with arbitrary height map $h(\boldsymbol{R_0})$ and given scattering amplitude. It has been used not only to prove well-known surface features, like atomic steps [86], but also to construct valuable morphology models of peculiar surfaces, as in the case of corrugated MnAs layers on GaAs(001) [89] or sub-surface line dislocations in magnetite thin films [90]. Moreover, the CTF algebra helps to figure out how to achieve the best performances from a cathode lens microscope. It is now clear the effect of the contrast aperture, which on one side deteriorates the image by acting as a low-pass filter and on the other side limits the acceptance angle and therefore the blurring induced by spherical aberration. The energy distribution of electrons coming, e.g., from the electron source, act together with the chromatic aberration, while the voltage and current instability can be modeled as an additional defocus. This knowledge has proven to be crucial in the case of aberration-corrected systems, where the lifetime of fully-corrected state has observed to be just a few minutes. After a correct estimation of every contribution, Schramm et al. [61] concluded that the stability of power supplies, the active damping of vibrations, good electromagnetic shielding and improved detectors are the crucial factors for maintaining the corrected state, and that more accurate monitor and correction systems must be developed to prolong its lifetime and make it usable for complex experiments.

### 3.1 Imaging Mode

#### 3.1.1 LEEM and LEEM-IV

The first operating mode of LEEM corresponds to the magnified image plane is displayed on the detector. Typically, the camera interconnects with the microscope software to capture single images or videos on varying the start voltage, the sample temperature, the lens settings and so on. In this way, several *in-situ* experiments and diagnostic procedures can be performed. In principle, the LEEM image is the intensity distribution of electrons on the image plane. The local intensity is then converted to a gray scale image, where the contrast depends of how the objects modify the reflected electron plane wave in phase and amplitude. In the following we show topical examples of how phase and amplitude objects are displayed in LEEM.

The simplest phase object giving contrast in LEEM is an atomic step on an elsewhere flat, crystalline surface. The uniform, regular distribution of atoms, as in the case of a terrace, gives no contrast in LEEM, since the electron beam is backscattered everywhere in the same way. Plane waves coming from two adjacent terraces have different phases, so that at the terrace edges the interference between them degrades the reflected intensity. **Figure 6a** shows a LEEM image of clean, stepped Si(111) surface with (7x7) reconstruction: monoatomic steps are displayed as dark lines with a faint, brighter decoration on one side [85]. This appearance is confirmed by simulation performed with CTF formalism [86]. In this case the surface height function $h(\boldsymbol{R_0})$ is expressed as a simple step function of height $a_0 = 0.31$ nm (**Fig. 6b**, green line). The simulated intensity line profile (**Fig. 6b**, blue line) displays a minimum in the vicinity of the step and a maximum on one side. In general, the presence of minima and maxima is related to the phase shift, i.e. the electron kinetic energy and the step height: For $\phi(\boldsymbol{R_0}) = 2n\pi$ the contrast is almost absent, while in the complete out-of-phase condition $\phi(\boldsymbol{R_0}) = (2n +$

1)$\pi$ the line profile is symmetric (no bright decoration). The asymmetric maximum is observed at the intermediate phase conditions and is most pronounced at $\phi(\boldsymbol{R_0}) = (2n + 1)\pi/2$. The relative position of maxima and minima can be inverted periodically as a function of the phase shift. CTF formalism can be readily extended to two spatial dimensions: **Figure 6c** shows the simulation of how an ideal Si(111) surface with monolayer-step-height circular and elliptical terraces (top) appear in LEEM for a given phase shift. It should be noticed that more complicated interference patterns could be produced when steps are close together, e.g., in the region highlighted with white arrows. The correct interpretation in such cases must pass through an extensive simulation of model surfaces in different conditions of focus, electron kinetic energy and morphology.

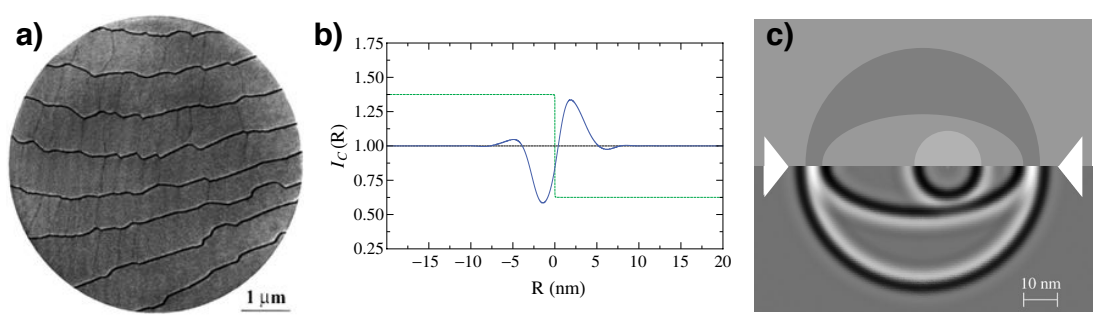

Figure 6: Imaging with phase contrast. (a) Underfocus LEEM image of monoatomic steps on the Si(111)-(7x7) surface. Imaging energy E0 = 45 eV. Reproduced from Ref. [85] with permission, copyright 1998 Elsevier. (b) Intensity line profile (blue) calculated for the superimposed step profile (dotted green line) with the inclusion of chromatic damping. (c) Plane view schematic (top) and simulated LEEM image (bottom) of an arrangement of terraces separated by a single atomic step. Bright regions of constructive interferences, where the steps are in close proximity, are arrowed. (b) and (c) reproduced from Ref. [86] with permission, copyright 2009 World Scientific.

The amplitude contrast is produced between two adjacent areas that have different scattering amplitude. This case is quite common during an experiment: any area with a different composition, stoichiometry, crystal structure and even surface reconstruction give amplitude contrast. The exact calculation from first principles of how crystalline surfaces diffract electron beams in LEEM is derived from kinematic and dynamic LEED theory developed already from late 1960s [83] and not discussed here in detail. An example of amplitude contrast is given in **Fig. 7a**, displaying LEEM image of Pt(111) surface covered with a graphene layer of variable thickness [91]. In this particular case, the contrast is given not only by changes in amplitude, as between monolayer and bilayer graphene, but also by the quantum size effect. The difference is more evident by looking at the IV characteristics obtained from a stack of LEEM images with increasing start voltage (**Fig. 7b**). The IV curve for ML graphene reflects the particular morphology of the system, where the carbon sheet rests 3.30 Å upon the Pt substrate. The other reflectivity curves appear quite similar to one another at kinetic energies above 20 eV, since the graphene thickness becomes bigger than the inelastic mean free path and no contribution from substrate atoms is present. Nonetheless, pronounced oscillations due to quantum size effects can be observed at lower kinetic energies (**Fig. 7c**). In particular, the number of minima of these oscillations scales with the number of layers. This characteristic has been observed not only on 2D materials, but also on epitaxial thin films on metal substrates, and therefore can be used as a universal fingerprint to estimate the film thickness. Moreover, one can plot the phase shift from Eq. 1 as a function of the energies at which interference maxima or minima are observed (**Fig. 7d**) and determine accurately the film thickness and the inner potential with a fit. In this particular case, the analysis confirms that graphene stacks thicker than 3 layers have identical layer separation to graphite.

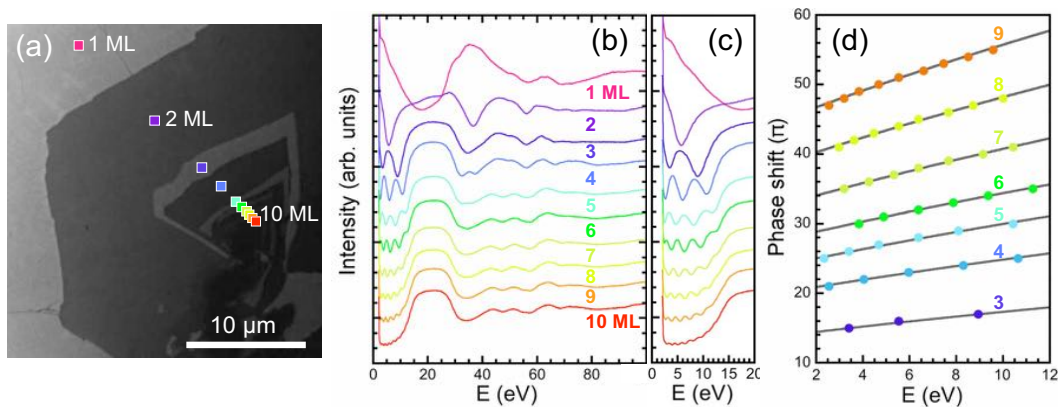

Figure 7: Imaging amplitude objects. (a) LEEM image (electron energy 4.4 eV) of a few-layer graphene stack nucleated at a boundary between rotationally misaligned ML graphene domains. Markers denote areas with coverage between 1 and 10 graphene layers. (b) IV characteristics obtained from a stack of LEEM images with electron energy from 2 to 100 eV at the locations marked in (a). (c) Higher magnification of the same data set at electron energies below 20 eV, showing fringes due to interference of electrons backscattered from the graphene surface and graphene/Pt interface. (d) Phase shifts for constructive and destructive interference [fringe maxima and minima in (c)] as a function of electron energy. Full lines are fits assuming free-electron like propagation. graphene. Reproduced from Ref. [91] with permission, copyright 2009 American Physical Society.

### 3.1.2 Brightfield and Darkfield LEEM

Up to now, we considered only the case in which incoming and reflected electron beams are perpendicular to the sample surface. Even with a perpendicular incoming beam, the outgoing electrons distribute over the solid angle to form a diffraction pattern in the backfocal plane of the objective lens. Then, the contrast aperture limits the acceptance angle in order to let pass only electrons emitted in the neighborhood of the zero-order diffraction spot. This configuration is called *brightfield* and is schematized in **Fig. 8a**. However, it would be interesting to build the LEEM image also with non-perpendicular electrons, i.e. with higher or fractional order diffraction spots; in this way the crystallographic information contained in the diffraction pattern can be transferred to the real space and generate a crystallographic map of the sample. Such case is called *darkfield* and can be achieved in different ways. The simplest way is (i) to move the contrast aperture and accept electrons with a non-zero emission angle, e.g., from a first-order diffraction spot (**Fig. 8b**). This method has the disadvantage that the selected electron trajectory is far from the optical axis and therefore the spherical aberrations may blur the image. A way to overcome this limitation is (ii) to incline the sample tilt by a certain angle $\alpha$ to let the selected diffraction feature with emission angle $2\alpha$ pass through the contrast aperture along the optical axis (**Fig. 8c**). In this case the incident electron beam is not perpendicular to the surface, so one should take into account the different atomic scattering factor $f(s)$, as it depends on the momentum transfer $s = k_{out} - k_{in}$. This approach does not need dedicate alignment of the lens system, but lacks of accuracy and reliability due to mechanical limitations of the sample manipulator. The most convenient method to produce darkfield LEEM is (iii) to leave the sample untouched and tilt the incoming electron beam with deflectors placed in the illumination column (**Fig. 8d**). It is optically equivalent to the previous case, but with the advantage that the sample holder and the lens system are untouched, and deflectors can be remotely controlled and accurately calibrated, enabling fast switch between brightfield and darkfield.

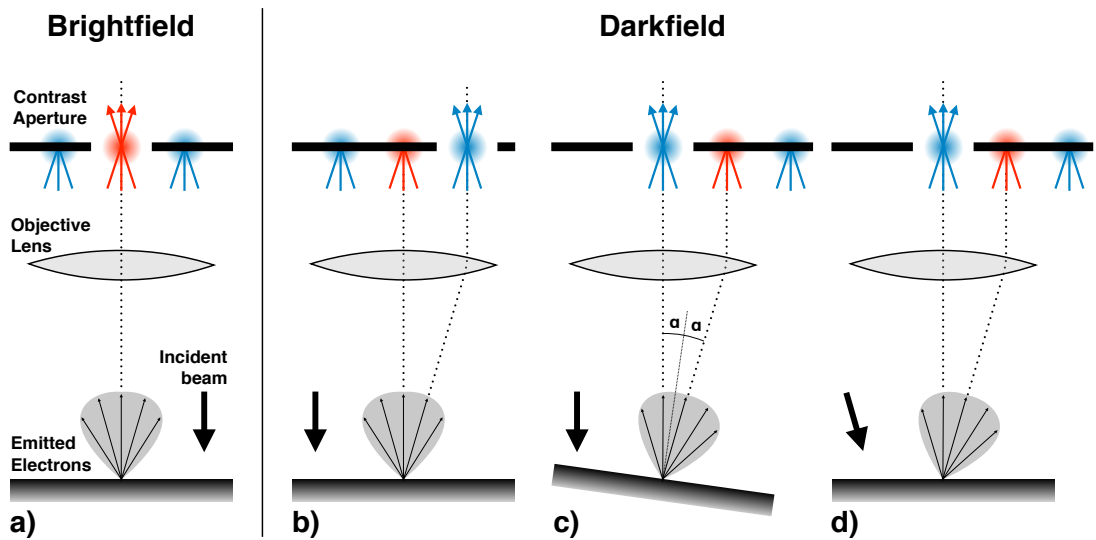

Figure 8: Scheme of brightfield (a) and darkfield LEEM operation. Darkfield can be performed (b) by displacing the contrast aperture, (c) tilting the sample, and (d) deflecting the incident electron beam. The zero-order diffraction spot is depicted in red, while the higher order ones are in blue.

The contrast mechanism in darkfield LEEM measurement adds substantial information on the crystal structure of the sample. Electrons forming a particular spot in the diffraction pattern are emitted only from areas with a certain crystal structure. Thus, a LEEM image produced with these electrons shows as bright the area from where they were emitted, and as dark the areas with another structure. By studying the area distribution in darkfield LEEM for several diffraction spots, one can reveal if a LEED pattern is produced from a single phase over the entire surface or is a superposition of two or more contributions. Images over a large field of view offer a direct measurement of the relative coverage of the phases. Moreover, domains with same crystal structure but rotated orientation can be distinguished. Even if the domains give the same LEED pattern geometry, the intensity of same-order spots differs, so resulting in an amplitude contrast among rotational domains. An example of both occurrences is given in **Fig. 9**. The system is a de-wetted

Fe$_3$O$_4$(111) thin film of thickness larger than 7 nm, grown on a Pt(111) surface [92]. The film holes do not expose a clean Pt surface, but are decorated by a single bilayer of FeO(111) [93, 94]. The corresponding LEED pattern, obtained with illumination over a large area, is a superposition of two distinct patterns: the (2x2) superstructure over the Fe$_3$O$_4$(111) spots (unit cell in green) and the moiré pattern of FeO(111) surrounding the Pt(111) spots. Brightfield LEEM (**Fig. 9a**) shows areas with different reflectivity, but at a first glance one cannot distinguish which is Fe$_3$O$_4$ and which is FeO/Pt. Darkfield LEEM performed with electrons from the moiré (**Fig. 9b**) shows as bright the FeO(111) areas, leaving the rest as dark. The contrast inverts when one of the Fe$_3$O$_4$(111) (2x2) spots is used, but while the FeO patches appear dark, only one rotational domain of Fe$_3$O$_4$ enlightens (**Fig. 9c**). The other rotational domain, rotated by 180°, emerges on darkfield LEEM by using the inequivalent (2x2) spot (**Fig. 9d**). It should be noticed that the FeO darkfield image corresponds to the sum of the two Fe$_3$O$_4$ darkfield images when the contrast is inverted, thus excluding the presence of a third crystalline phase. Moreover, the amplitude contrast between rotational domains is achieved only at some kinetic energies, for which two inequivalent LEED spots with same order have a different intensity. Other energies can eliminate or invert the contrast.

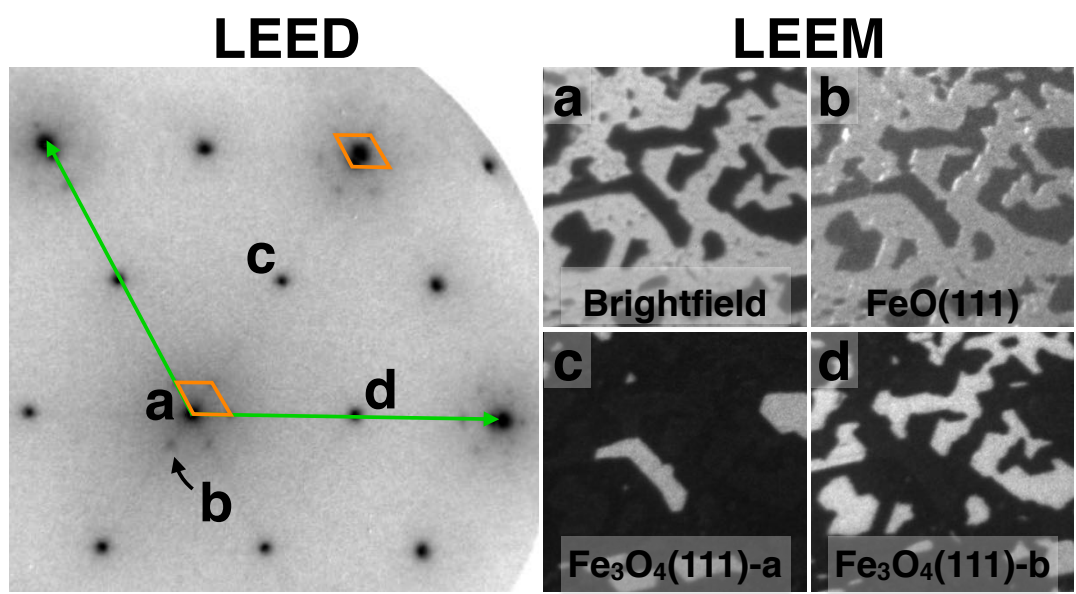

Figure 9: LEED (left) and LEEM (right) images of a strongly dewetted Fe$_3$O$_4$(111) thin film. In LEED ($E_0$ = 88 eV) the reciprocal vectors of Fe$_3$O$_4$(111) unit cell and FeO(111) moiré pattern are highlighted in green and orange, respectively. The labels in LEED mark the selected diffraction spots used for LEEM ($E_0$ = 24 eV): (a) brightfield, (b) darkfield with FeO(111) moiré, (c) and (d) darkfield with (2x2) inequivalent spots of Fe$_3$O$_4$(111). Reproduced from Ref. [92] with permission, copyright 2012 American Physical Society.

### 3.1.3 Mirror Electron Microscopy

It has been shown that in LEEM the reflectivity of electrons changes with the momentum transfer *s* occurring during backscattering. Such reflectivity is always less than unitary, i.e. part of the electrons is lost due to inelastic scattering, bulk absorption, surface and quantum effects. The only way to achieve total reflection is to decrease the start voltage until all electrons are reflected above the sample surface, turning the cathode immersion lens into an electrostatic mirror. However, the equipotential surface in front of the specimen is still influenced by field inhomogeneities determined by the surface morphology, work function changes, contact potentials and magnetic fields. As electrons decrease speed and reverse direction, their trajectories are deviated by such perturbations, thus giving contrast to the electron image. This imaging technique is called Mirror Electron Microscopy (MEM) and has the advantage to probe surfaces without direct impact, giving access to non-conductive specimens and imaging phenomena in a non-perturbative way. The contrast mechanism in MEM has been discussed and modelized over the years [95–97], in order to extract quantitative information regarding morphology and microfields. Although the algebra is in some cases quite similar to the CTF approach described in Sect. 3.0.2, it will not be discussed here. In general, the lateral resolution in MEM is worse than standard LEEM imaging on the same surface, ranging around several tens of nm. Nonetheless, the high sensitivity to height variations and equivalent surface potentials gives a remarkable depth resolution of about 1 nm. Like in LEEM, the intensity line profile of features in different focus conditions can be simulated and reverted to quantitative real-space models. I(V) spectra of different areas through the LEEM-MEM threshold can be used to extrapolate a map of the local potential, owed to work function changes, charge states or application of external fields.

### 3.1.4 Spin Polarized LEEM

Spin is a degree of freedom of the incident electron beam that can be used to achieve imaging of magnetic states of the specimen surface. As shown in Sect. 2.2.3, spin-polarized sources such as III-V semiconductor photocathodes can provide beams with high degree of polarization $\boldsymbol{P}$. In SPLEEM, the usual image contrast is augmented by magnetic contrast generated by the exchange interaction between incident spin-polarized electrons and spin-polarized electrons in the magnetic material [98–101]. This exchange contribution to the scattering is proportional to $\boldsymbol{P} \cdot \boldsymbol{M}$, being $\boldsymbol{M}$ the magnetization vector of the target material. In a magnetic material $\boldsymbol{M}$ results from the difference between the number of electrons with parallel and antiparallel spin contained in the occupied states of majority and minority bands, respectively (**Fig. 10a**) [102]. The two electronic populations produce non-equivalent exchange-correlation potentials, so that electron beams with different polarization are scattered differently. Moreover, the minority spin band offers more unoccupied states for an inelastic event, thus minority electrons are more effectively scattered than majority electrons and the IMFP decreases. This leads to a larger reflectivity for majority electrons. The intensity difference between parallel ($I_{\uparrow\uparrow}$) and antiparallel ($I_{\uparrow\downarrow}$) configurations, normalized to the sum of the intensities, i.e.

$$A = \frac{I_{1\uparrow} - I_{1\downarrow}}{I_{1\uparrow} + I_{1\downarrow}} \quad ,$$

is called exchange asymmetry and is proportional to **P · M**, weighted with the damping caused by the different IMFP. It should be noticed that the difference at the numerator cleans the resulting image from non-magnetic diffraction and topological features resulting from conventional LEEM imaging, leaving only contrast from magnetic features. The effect of spin on exchange correlation potential and IMFP decreases rapidly as the kinetic energy of incident electrons increases. For this reason, the best magnetic contrast in SPLEEM is obtained at energy typically below 20 eV.

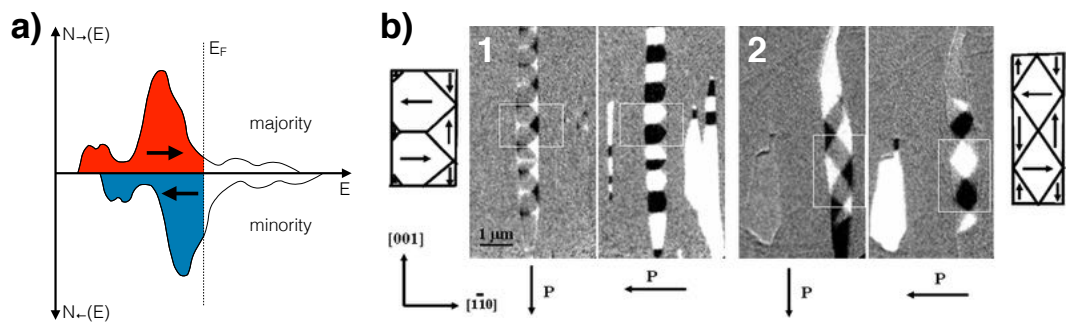

Figure 10: (a) Density of states in a ferromagnetic metal. Due to the spin interaction, the electronic band can be conceived as a superposition of a majority (red) and minority (blue) population. (b) Domain structure of epitaxial Fe ribbon crystals on W(110). The image pairs were taken with the polarization vector of the electrons parallel to the [001] (1) and [1$\underline{1}$0] (2) directions, respectively. The magnetization distribution in the marked regions is indicated on the sides. Reproduced from Ref. [103] with permission, copyright 2006 John Wiley & Sons.

The photocathode electron gun delivers electron beams with a fixed spin polarization vector that can be eventually flipped by changing the versus of the circular polarized light. The spin polarization vector can be subsequently changed with a spin manipulator, where electrostatic and magnetic deflectors and a magnetic rotator lens give three degrees of freedom on the spin orientation. This allows complete characterization of the sample magnetization direction in both in- and out-of-plane geometry, and tilted directions in between. An example of SPLEEM asymmetry images with different polarization orientation is given in **Fig. 10b**. Here, two epitaxial Fe ribbons produced by deposition of 5 ML of Fe on W(110) surface and annealing at 650 K are displayed with the polarization vector in-plane and parallel to [001] (1) and [1$\underline{1}$0] (2) crystallographic direction, respectively [103]. The magnetic state of the ribbons is primarily determined by the interplay between exchange and stray-field energy, which prefers magnetization along the [001] axis, and the surface/magnetoelastic energy, whose minimization produces states with magnetization along [1$\underline{1}$0]. The consequent multidomain state can be extracted by the intensity pattern in the SPLEEM asymmetry image: brightest areas have a parallel magnetization vector, darkest have an antiparallel one, while neutral grey are oriented perpendicularly. By combining images with different polarization vectors one can construct a consistent domain model, as presented on the sides for the marked regions.

SPLEEM has been used to address phenomena such as domain wall structures in thin magnetic films, micromagnetic configurations in surface-supported nanostructures, spin reorientation transition, magnetic coupling in multilayers, phase transitions and finite size effects. Its application has several advantages, such as real-time observation and possibility to combine crystallographic and magnetic information. The surface sensitivity limits its usefulness to samples prepared *in situ* or grown elsewhere and protected by a removable capping layer. The main disadvantage in the use of SPLEEM is its strong sensitivity to applied magnetic fields, which distress the trajectory of electrons and degrade the image quality. Modest fields of few hundred gauss can be applied only in the surface normal direction, so that the Lorentz force is geometrically minimized. This limitation affects important fields of research, such as dynamics on domain walls and exotic magnetic states of matter.

### 3.1.5 Electron Energy Loss Microscopy

LEEM systems equipped with an energy filter have the possibility to use inelastic electrons for imaging with opportune detuning of the energy analyzer [77, 104]. Electrons can lose some kinetic energy during the scattering process and the energy distribution of all inelastically scattered electrons provides information about the local physical and chemical properties of the specimen. The low-loss region (< 50 eV) of this energy spectrum contains valuable information about the band structure and the dielectric properties of the material, e.g., electron-phonon interaction, band gaps and surface plasmons [105, 106]. Such inelastic electrons can pass through the energy analyzer with optimal trajectory if a supplementary bias is applied. The usual slit at the exit plane selects only electron with a certain energy loss. EELM images have typically very low intensity and contrast, but can be used to display surface distribution of plasmons and to distinguish between surface areas with different phononic and plasmonic properties [107]. For example, this is the case when graphene (Gr) and hexagonal BN (h-BN) patches rest one aside the other upon a surface [108]. **Fig. 11a** shows a EELM image of adjacent Gr and h-BN flakes grown on Pt(111) surface from a single molecular precursor, dimethylamine borane (DMAB). The simultaneous presence of B, C and N atoms obtained from dissociation of DMAB at 1000 K is a very efficient way to obtain a continuous, almost free-standing layer mostly made of Gr and h-BN, with only a low percentage of impurities. Gr and h-BN flakes have a different plasmon energy loss and therefore display a contrast for particular electron loss energies. Local integration over a stack of EELM images allows the collection of size-selected electron energy loss spectra (**Fig. 11b**). It is shown that in the bright areas in EELM the collective excitation of the electrons is found at 6.5 eV, whereas in the dark areas a peak centered at 7.7 eV is found. These spectroscopic features are assigned to π-plasmon energy loss in slightly doped Gr and h-BN, respectively [109, 110].

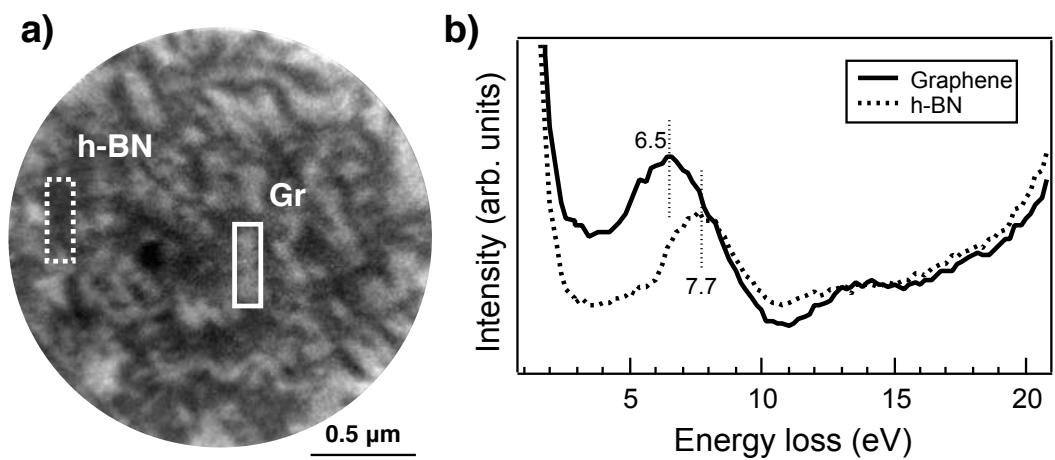

Figure 11: (a) Electron Energy Loss Microscopy image of graphene and h-BN coplanar flakes on Pt(111) collected with electron energy $E_0$ = 32 eV and a loss of 6.5 eV. Graphene patches appear brighter than h-BN due to the plasmonic excitation. (b) Electron Energy Loss spectra of graphene and h-BN extracted from a stack of EELM images at different energy losses. The spectrum measured in region Gr shows a π-plasmon loss at 6.5 eV; the spectrum measured in region h-BN shows a π-plasmon loss at 7.7 eV. Reproduced from Ref. [108] with permission, copyright 2015 Wiley-VCH Verlag.

The use of energy loss microscopy in LEEM systems can provide additional information on the dielectric nature of surfaces and thin films. However, its accuracy cannot be compared to a dedicated apparatus. In LEEM the monochromaticity of the incident electron beam, few tenths of eV, is not enough to resolve vibrational states of molecules and adatoms. Despite this aspect, EELM is ideal to characterize inhomogeneous surfaces, showing the lateral extent of every species with different plasmonic signature with a resolution of some tens of nm.

## 3.2 Diffraction mode

### 3.2.1 µ-LEED

The second operating mode in LEEM is the so-called diffraction mode, i.e. when the backfocal plane of the objective lens is displaced on the detector. This mode gives access to the angular distribution of backscattered electrons, which forms a diffraction pattern in case of crystalline surfaces. The use of LEEM systems for diffraction studies has many advantages respect to standard LEED optics:

- The operation conditions of the electron gun and the illumination angle are fixed, while the kinetic energy at the interaction is governed by the start voltage. This ensures beam stability and constant current, even for dynamic measurements.
- The backfocal plane is displayed for electrons traveling at a kinetic energy $E$, independently of their takeoff energy $E_0$. This means that the displayed reciprocal space has the same lateral extent for every start voltage, so the diffraction spots do not move during an energy scan. The calibration of the reciprocal space can be calculated by looking at the linear expansion of the Ewald sphere with increasing start voltage, or through the position of the diffraction spots for a known surface, e.g., graphene, Si(111)-7x7 or oxidized W(110).
- The probed region can be inspected in LEEM and selected by placing an opportune aperture in the image plane. Commercial LEEM systems can reduce the illuminated area to a diameter of 250 nm [30].
- The electrons can be filtered in energy and the background of secondary electrons can be removed.
- The zero-order diffraction spot can be easily detected, as the magnetic beam splitter separates the incoming and outgoing electron beams.

The collection of LEED measurement from a selected region is often referred to as micro-LEED or µ-LEED. **Fig. 12** shows how microscopy and diffraction can be combined to obtain structural information on surfaces and nanostructures under particular conditions. The LEEM image in **Fig. 12a** presents cerium oxide microparticles grown on a Ru(0001) surface saturated with oxygen [111, 112]. This system is a model catalyst, used to study the interplay between oxide and metal under reaction conditions. A 500 nm wide illumination aperture can be introduced and placed on a large $CeO_2$ particle (as indicated by the red circle), so that electrons are backscattered from only this single object. The transfer lens setup is then changed to display the LEED pattern. The real-time observation of the diffraction pattern was used to investigate how the reduction of ceria particles influences their atomic surface structure. LEED patterns were acquired in real time while dosing up to 4800 L of molecular hydrogen at a substrate temperature of 700 K (**Fig. 12b**). Before hydrogen exposure, only the (1x1) integer spots of $CeO_2$ are visible. After dosing 500 L of $H_2$ at $5 \times 10^{-7}$ mbar, additional spots emerge as a consequence of the local ordering of oxygen vacancies induced by $H_2$ dissociation and surface reduction. At this moment, the superstructure spots exhibit a periodicity of 2.6 respect to the integer spots of $CeO_2$. Further dose of $H_2$ at higher pressure leads to larger periodicities in the diffraction pattern, notably (3x3) at 1900 L and (4x4) at 4800 L, as well as slight in-plane lattice expansion, detectable from the contraction of first order spots. The structural changes observed in LEED, together with other LEEM analysis not shown here [111], helped the authors to conclude that under reducing conditions three stable phases of reduced ceria exist, which coexist for intermediate oxidation states.

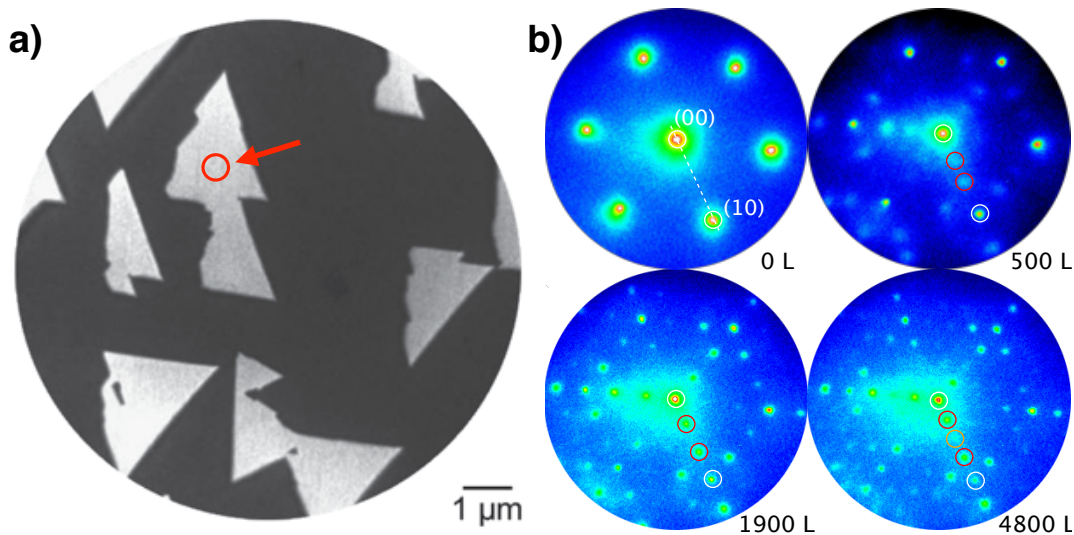

Figure 12: (a) LEEM image recorded at 16.3 eV of ceria microparticles (bright) on the Ru(0001) support (dark). The open circle highlighted with an arrow illustrates the electron beam spot size and position during μ-LEED. (b) μ-LEED image series obtained during reduction of a single ceria microparticle in hydrogen at 700 K. White circles indicate the reflections of CeO$_2$. Red and orange circles indicate the positions of the superstructure spots. Reproduced from Ref. [111] with permission, copyright 2015 Wiley-VCH Verlag.

### 3.2.2 Spot Profile Analysis and LEED-IV

The real-time observation of energy-filtered, stable LEED pattern is a valuable tool to investigate inhomogeneous surfaces under changing conditions. The quality of LEED images taken in LEEM is such that other analysis methods are made available. For example, the intensity line profile of a particular diffraction spot along a selected crystallographic direction can reveal many details on the surface morphology and roughness, such as step distribution, presence of defects and ordered superstructures [113]. The sampling frequency of the profile is determined by the magnification of the backfocal plane and the number of pixel of the detector. The sharpness limit for LEED spots is determined by the transfer width of the electron beam at the surface and depends on the type of electron source and other instrumental effects of the LEEM apparatus.

**Figure 13a** gives an example of a LEED Spot Profile Analysis (SPALEED) performed with LEEM optics on a Fe$_3$O$_4$(111) thin film grown on Pt(111) substrate [92]. Magnetite films are used both as a model catalyst and a support for catalitically active nanoparticles. Its surface termination, deeply connected with its functional and catalytic properties, changes with the preparation conditions. In this case the film was grown with subsequent cycles of Fe deposition and oxidation at 900 K. After the last oxidation performed at 1000 K, if the sample is cooled in oxygen atmosphere, the zero-order diffraction spot presents a shoulder-like broadening. Such broadening disappears after flashing at 900 K in UHV. The (0,0) spot profile can be fitted with a superposition of a Gaussian peak, accounting for the instrumental broadening, and a shoulder that can be described as a sum of three Lorentz$_{3/2}$-like functions of different half widths [114].

$$I(\mathbf{k}) = I_{Gauss}(\mathbf{k}) + I^{(1)}_{Lor3/2}(\mathbf{k}) + I^{(2)}_{Lor3/2}(\mathbf{k}) + I^{(3)}_{Lor3/2}(\mathbf{k})$$

The prominent broadening is described by the first part, $I^{(1)}_{Lor3/2}(\mathbf{k})$. The weakly modulated background, ascribable to small clusters or adsorbates on the surface, is described by $I^{(2)}_{Lor3/2}(\mathbf{k})$ with a full width half maximum (FWHM) as large as the first Brillouin zone. The third component, $I^{(3)}_{Lor3/2}(\mathbf{k})$, has a FWHM slightly larger than the Gaussian one and can be related to the presence of atomic steps. The spot profile has then been collected over an energy range between 40 and 200 eV, in order to highlight the changes in the relative intensities of the components. It has been found that the FWHM of the components increased linearly with the perpendicular component of $\mathbf{k}$, indication that the surface has a mosaic structure with a calculated angular spread of 0.2°. Moreover, the ratio $G$ between integral intensities of $I_{Gauss}(\mathbf{k})$ and $I^{(3)}_{Lor3/2}(\mathbf{k})$ spot components,

$$G = I_{Gauss}/(I_{Gauss} + I^{(3)}_{Lor3/2}) \quad ,$$

revealed a periodic exchange between the two intensities with period linearly related to the perpendicular component of $\mathbf{k}$. Such behavior is consequent to the periodic constructive and destructive interference between two adjacent terraces separated by an atomic step [113]. By fitting this periodicity one can calculate the step height: in this case, it was found to be 4.79±0.09 Å, in fair agreement with the height of the magnetite unit cell (4.84 Å).

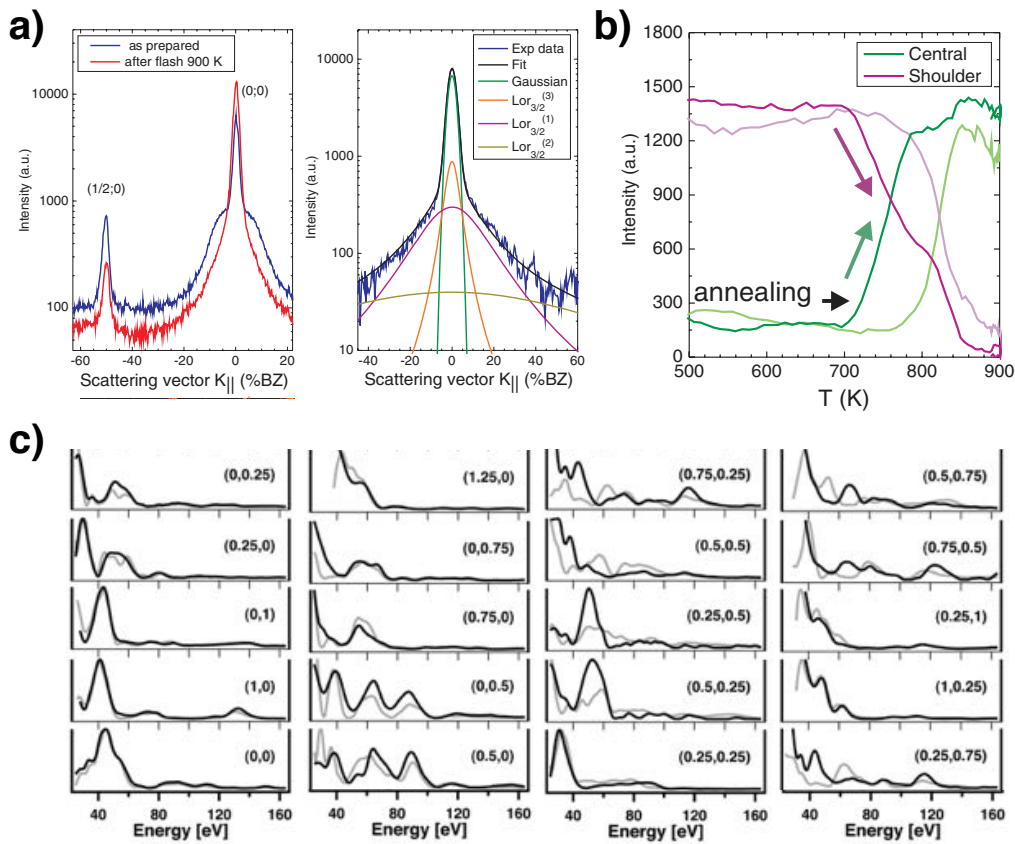

**Figure 13:** (a) Spot Profile Analysis of $Fe_3O_4$(111) LEED pattern along the (1,0) vector for the as-prepared surface and after a final annealing at 900 K. The preparation conditions of $Fe_3O_4$(111) thin film are described in the main text. The fit of the (0,0) LEED spot profile on the right shows a narrow central Gaussian peak in green, two Lorentz$_{3/2}$ peaks in orange and purple for the shoulderlike broadening and a very broad Lorentz$_{3/2}$ peak in dark yellow. (b) Integral intensity of the central Gaussian and the shoulder during cooling in oxidation conditions. The formation of surface inhomogeneities is influenced by the cooling rate, ~ 4 K/s for dark lines and ~ 1 K/s for light lines. (a) and (b) reproduced from Ref. [92] with permission, copyright 2012 American Physical Society. (c) Comparison of experimental IV curves (black) taken with μ-LEED and best-fit calculated curves (gray) for (4x4) diffraction structure of oxidized Ag(111). Reproduced from Ref. [115] with permission, copyright 2007 American Institute of Physics.

The nature of the objects giving the shoulder-like broadening was then clarified with real time acquisition of LEED pattern in dynamic conditions. **Fig. 13b** shows the intensities of the Gaussian and the first Lorentz$_{3/2}$-like spot components during the formation at the cooling in oxygen atmosphere for two different cooling rates. The broadening in the zero-order diffraction spot and its behavior during cooling in oxidation conditions suggest that the prepared surface is roughened by oxygen-related objects smaller than the lateral resolution in LEEM [92]. Such objects can expose atoms with different coordination and charge states respect to the ideal surface and therefore influence the catalytic behavior of the system [116].

Like in the case of LEEM, the collection of LEED patterns over a broad energy range gives access to structural information. The intensity modulation of diffraction peaks can be simulated with full dynamic calculations. The calculation packages available nowadays allow the comparison of calculated and experimental IV curves to find the most probable atomic configuration of the surface. The acquisition of LEED-IV spectra in a LEEM system starts with the collection of a stack of LEED images with variable start voltage. The intensity of a single peak can be extracted with proper fitting of the line profile, as in the case of SPALEED, or with direct integration over a rounded region of interest of the image. In both cases, a background subtraction is required. To perform a reliable dynamic analysis, the dataset, calculated as the sum of the energy widths for every inequivalent diffraction spot, must be in the range of thousands of eV [83, 117]. The IV curves must then be smoothed by a convolution with a Lorentzian curve of width 1-2 eV to reduce the noise [118]. The first successful IV analysis performed with LEEM optics regards the (4x4) superstructure created by surface oxidation of Ag(111) [115]. The comparison of experimental IV curves and calculated ones for the best fit model is shown in **Fig. 13c**. In this experiment, the dataset was compared with one recorded with a conventional LEED system. It is shown that LEED acquisition with LEEM optics offers a much better signal-to-noise ratio (some faint diffraction spot were visible in LEED/LEEM and undistinguishable from background in conventional LEED) and a larger energy range for every spot.

The combination of LEED and LEEM gives another advantage respect to conventional LEED optics [90, 92]. In the latter one, the illuminated area is very large, in the order of several hundreds of μm. In case of a surface with rotational domains, some inequivalent spots from different domains of unknown relative abundance can superpose, thus forcing to average spectra of spots of the same diffraction order and making impossible the detection of separated spectra. In LEED/LEEM systems the active selection of the illuminated area can solve this problem. The probed region can be inspected with darkfield LEEM using an inequivalent diffraction spot and the relative abundance $p$ of the domains can be calculated. Then, the disentangled IV spectra can be extracted as follows. Assuming that the experimental LEED pattern is a weighted, incoherent superposition of the rotated LEED patterns, in case of two rotational domains one can write that

$$I_{total}(\boldsymbol{k}_{xy}, k_z) = (1-p)I_+(\boldsymbol{k}_{xy}, k_z) + pI_-(\boldsymbol{k}_{xy}, k_z),$$

where $I_+(\boldsymbol{k})$ and $I_-(\boldsymbol{k})$ are the intensity of two inequivalent spots of the same order, produced by two domains with abundance $(1-p)$ and $p$, respectively. The symmetry condition between rotational domains imposes an equivalency between same-order spots. In case of two domains with 180° symmetry,

$$I_+(\boldsymbol{k}_{xy}, k_z) = I_-(-\boldsymbol{k}_{xy}, k_z)$$

As a consequence, one can separate the two contributions:

$$I_+(\boldsymbol{k}_{xy}, k_z) = \frac{1-p}{1-2p} I_{total}(\boldsymbol{k}_{xy}, k_z) - \frac{p}{1-2p} I_{total}(-\boldsymbol{k}_{xy}, k_z)$$

With this method, IV spectra of inequivalent spots become accessible, thus improving the dataset quality and the reliability of its dynamic analysis.

### 3.3 Spectroscopy mode

#### 3.3.1 µ-EELS

If the LEEM apparatus is equipped with an energy filter, its dispersive plane can be displayed on the detector and spectroscopy over a selected area can be performed. The energy spectrum of reflected low-energy electrons presents a very intense peak of elastically scattered electrons, a weaker tail due to inelastic events and the secondary yield. In a region close to the elastic peak, one can perform Electron Energy Loss Spectroscopy from a micron-sized area (µ-EELS) and detect the electrons that interact with the specimen via creation of plasmons and phonons, in the same fashion of **Fig. 11b**. Although the direct measurement of the dispersive plane offers a better energy resolution and signal-to-noise ratio than the spectrum extrapolation from EELM images, the performance is still too low compared to a dedicated EELS apparatus [105, 106]. Other loss features at higher energies, such as ionization edges of core level electrons, are poorly visible from an apparatus optimized for low-energy electrons. For these reason, the spectroscopy mode in a LEEM system is used marginally and only as a complement to other investigation techniques. Its counterpart with emitted electrons is much more scientifically valuable, therefore a more complete technical review of the mode can be found on Section 4.3.

## 4. PhotoEmission Electron Microscopy

PEEM is equipped with a photon source that illuminates the specimen with ultraviolet and X-ray photons. Electrons emitted via photoelectric effect are collected by the cathode immersion lens and used as information carriers. The image formed on the detector can be acquired in real time like in LEEM, but the acquisition time of a single frame can vary considerably as a consequence of the electron beam intensity: while video-rate is still possible with secondary electrons for intermediate magnification, the collection of a single XPEEM image with core-level electrons requires minutes. The nature of image contrast in PEEM is different from the case of LEEM and is governed mainly by the physics behind the photoemission process and the chemical state of the probed matter. Here only a general discussion will be presented, followed by a more detailed examination of space charge phenomena that affect PEEM with pulsed light sources.

#### 4.0.1 Basic Image Contrast

The main difference in image formation between PEEM and LEEM is the electron coherence. The photoemitted electrons are in general incoherent in time and space, although for delocalized valence-band electrons some coherence can be seen in reciprocal space. As a consequence, the intensity distribution of the image is the convolution of the intensities of the phase object and the contrast transfer function (while in LEEM, i.e. under coherent conditions, the intensity is the square modulus of the convolution) [88]. By following the definition given in Sects. 3.0.1 and 3.0.2, one has that

$$I(\mathbf{R}) = |\psi_{obj}(\mathbf{R})|^2 \otimes |T_{inco}(\mathbf{R})|^2$$

The incoherent process modifies the definition of the contrast transfer function. In the coherent case the chromatic damping envelope $\mathcal{E}(\mathbf{q}, \Delta E)$ comes from an integration over the weighted contribution of the different energies within the Gaussian energy distribution of the incoming beam. For PEEM, the temporal incoherence imposes a different algebra. Schramm et al. [88] have modeled the square modulus of the PSF $|T_{inco}(\mathbf{R})|^2$ as the Fourier transform of the square modulus of the CTF for the monochromatic case weighted over the energy distribution $N(E)$,

$$|T_{inco}(\mathbf{R})|^2 = \mathcal{F}^{-1}\left[\int_{-\infty}^{\infty} |H_{mono}(\mathbf{q}, E)|^2 N(E) dE\right]$$

The electron incoherence implies that only amplitude objects give contrast: in PEEM phase objects like atomic steps and other morphologic features are generally invisible. **Figure 14a** reports one of the first images taken in 1934 with an improved version of the first Brüche's PEEM [119, 120] and shows a polycrystalline Pt foil annealed at high temperature. The contrast is given by the work function difference among the facets and the grain boundaries. The lateral resolution for amplitude objects depends on the kinetic energy of the electrons and their energy and angular distribution. In UVPEEM a standard Hg Short-Arc lamp induces photoemission of valence band electrons with a $\cos\theta$ angular distribution over an energy window $\Delta E$ larger than 1 eV. In this case, it has been calculated that the best lateral resolution should be about 7 nm for non-corrected instruments and less than 4 nm for the aberration-corrected case [88]. Similar values can be expected also in energy-filtered XPEEM for low start voltages [49], while for higher takeoff energies the performance goes worse. Such limits were partially reached when PEEM systems with UHV technology were made available. Gertrude Rempfer and coworkers demonstrated in the late 1980s a lateral resolution of 7 nm with a non-aberration-corrected UVPEEM [33]. **Figure 14b** shows a photoelectron micrograph of colloidal silver particles taken with her instrument. Aberration-corrected instruments lowered the limit for UVPEEM to about 5 nm.

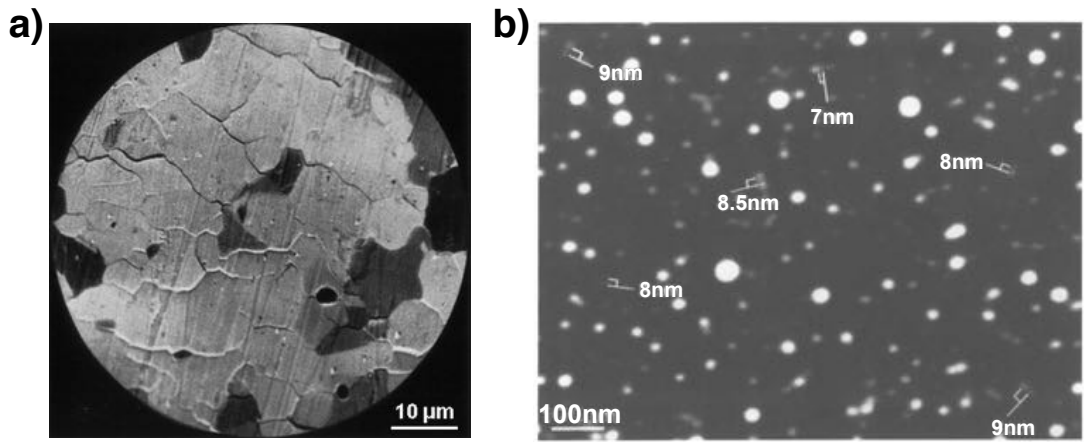

**Figure 14: (a) Brüche's PEEM image of a polycrystalline Pt foil after annealing at high temperature. Reproduced from Ref. [120]. (b) PEEM image of colloidal silver demonstrating 7 nm resolution. Reproduced from Ref. [33] with permission, copyright 1992 Elsevier.**

### 4.0.2 Space charge effects

The lateral resolution in XPEEM is a particular case that deserves a dedicated discussion. The use of monochromatic synchrotron radiation to excite core-level electrons and employ them for imaging has brought huge advantages in terms of flux and versatility, but its pulsed time structure has a worsening effect on both lateral and energy resolution. When a pulsed light beam illuminates the sample, the photoemitted electrons are packed in a small volume. During the flight the cloud density is such that electrons experience reciprocal Coulomb repulsion, thus degradating the transported information [121]. This effect, called space charge, has been broadly studied for electron sources and influences both the energy distribution (Börsch effect [122]) and the trajectory displacement (Löffer effect [123]). Fewer studies addressed the problem in XPS [124] and specifically in PEEM, not only using synchrotron radiation [62, 78], but also fs-laser [75, 125].

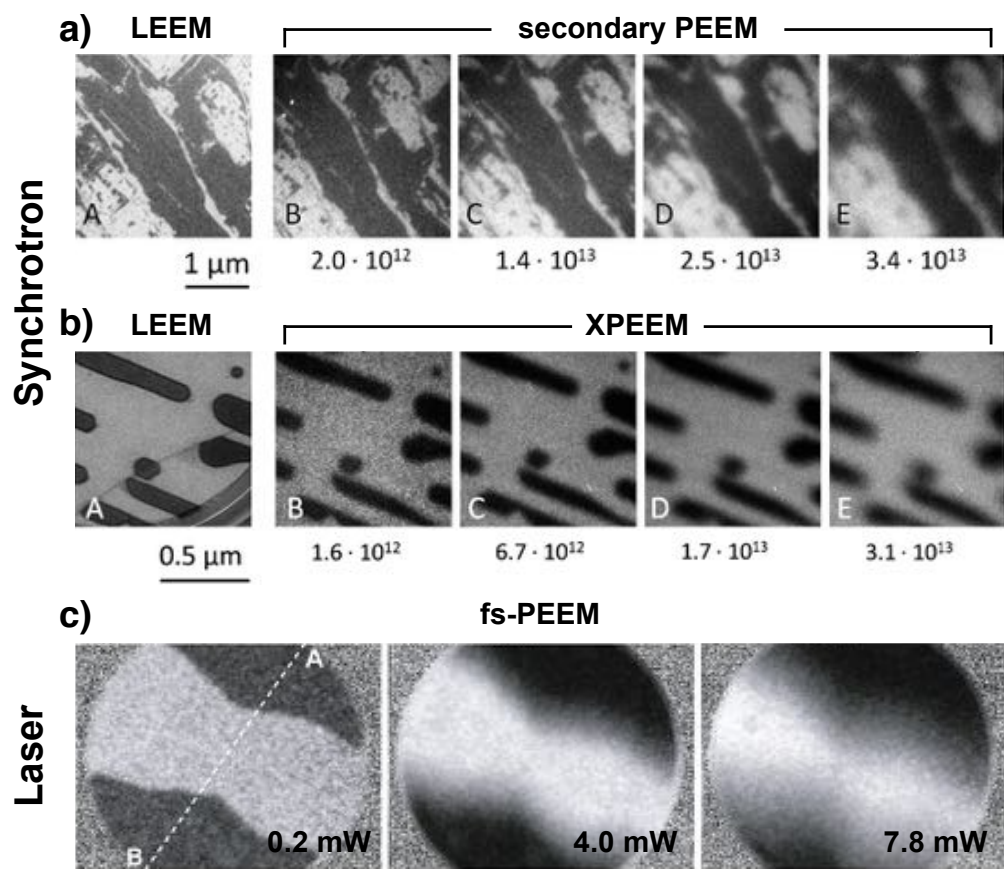

Figure 15: (a) LEEM and secondary XPEEM images of one ML thick Au islands on Ir(001). $E_0$ = 4 eV. (b) LEEM and W 4f7/2 XPEEM images of thick Fe islands on W(110). $E_0$ = 147 eV. Both series of XPEEM images illustrate the degradation of the microscope lateral resolution with increasing photon flux, which is indicated below each image (units: photons/s). Photon energy: 182 eV. The operating conditions of the microscope were the same in LEEM and XPEEM. Reproduced from Ref. [78] with permission, copyright 2010 Elsevier. (c) Magnetic circular dichroism fs-PEEM images at increasing laser power of 12 ML Ni deposited on Cu(001) and covered by Cs to reduce the workfunction. The photon source was a femtosecond laser with a photon energy of 3.1 eV. The lateral resolution was evaluated along the A-B cross section. Reproduced from Ref. [75] with permission, copyright 2009 IOP Science.

In the specific case of synchrotron radiation, the photon beam used for XPEEM delivers typically ~ $10^{13}$ photons/s of variable energy on a small area of few tens of µm². The photon flux is not constant in time, but is organized in pulses. Every pulse has duration of few tens of ps and hosts about 20000 photons. The number of photoelectrons emitted during a single pulse depends on the total photoionization quantum yield: in the case of photoemission from transition metals, this can be estimated as about 400-600 electrons. However, just a few of them are core-level electrons used directly in XPEEM, while the overwhelming majority (~ 97%) are secondary electrons emitted by inelastic events and filtered away in a second moment by the energy analyzer. In order to understand the magnitude of the space charge effect, Locatelli et al. [78] showed LEEM and XPEEM images of the same surface at different photon flux. In LEEM the continuous flow of incident electrons ensures a photocurrent density three orders of magnitude lower than the one in XPEEM, producing a space-charge free image. The latter was then compared to secondary electron ($E_0$ = 4 eV) XPEEM images acquired with the same alignment and configuration of the microscope, but for different photon fluxes (**Fig. 15a**). The lateral resolution in secondary XPEEM increases linearly with the photon flux, from 35 nm with $2.0 \times 10^{12}$ photons/s to 180 nm with $5.5 \times 10^{13}$ photons/s. Such degradation is visible also in imaging with core level electrons. **Figure 15b** shows LEEM and XPEEM images of the same surface, using backscattered electrons and W $4f_{7/2}$ core level electrons at a kinetic energy of 147 eV, respectively. Again, the XPEEM image blurs as the photon flux rises. The lateral resolution ranges from 31 nm with $2.0 \times 10^{12}$ photons/s to 82 nm with $5.9 \times 10^{13}$ photons/s, a degradation smaller than the case of secondary PEEM. Analogous effect can be seen on the energy resolution. Locatelli et al. reported a broadening of the Fermi level on a Au(100)-*hex* surface in a series of spectroscopy measurements with increasing photon flux. At high fluences, the broadening induced by space charge effect dominated over the contribution of the energy analyzer.

Space charge effects occur also when PEEM is performed with femtosecond laser. Nakagawa et al. [75] proved that the laser light from a Ti:sapphire laser with pulses of 100 fs produces blurred PEEM images with worse resolution than using a CW laser. **Figure 15c** shows images of the same surface acquired with photons of energy 3.1 eV and increasing incident laser power, from 0.2 to 7.8 mW. In this case, the photon flux is much higher than the one with synchrotron light, $8 \times 10^{10}$ photons per pulse. Again, the lateral resolution is inversely proportional to the laser power, so one is forced to greatly reduce the power output and increase the integration time to obtain good lateral resolution. Similar results were found by Buckanie et al. [125] using a Ti:sapphire laser with comparable pulse length and lower photon flux. Such evidences demonstrate how space charge effects are the biggest restriction for the implementation of pump-probe experiments in PEEM, forcing users to work at low power output, with high repetition rates and long acquisition times.

In order to overcome the resolution limitation induced by space charge effect, it is important to understand more in detail the evolution of the electron beam from the emission to the detection. From the observation, it is clear that the space charge effect is determined by the number of photons per time per emitting surface. The X-ray photoemission with pulsed light produces a large majority of low energy secondary photoelectrons and a small amount of core level electrons with high energy, all taking off from the sample surface in a time range equal to the pulse length. Then, the high voltage between sample and objective lens accelerates the electron cloud. The lens system of the microscope induces several crossovers of the electron beam along the path, i.e. where the diffraction and image planes are placed. The first crossover is located in the backfocal plane of the objective lens, while the second is the first intermediate image plane. The magnification induced by the objective lens implies that the lateral size of the crossovers is larger than the initial size of the electron cloud at the takeoff. Therefore, no special point with significantly large space charge density may cause the local image blurring, i.e. the blurring occurs along the entire electron path. Moreover, simple kinematic calculation show that the strong overlap between secondary and core level electrons within one single pulse decays with time due to their different flight speed. For the SMART microscope, core level electrons with a take off energy of 100 eV are separated from the secondary electron bunch after 70 cm of flight. Behind this point the electron-electron interaction can be neglected for XPEEM measurements. The best strategy to reduce the space charge blurring is then to remove non-necessary electrons as soon as possible along the electron path. In particular, Schmidt et al. demonstrated that the insertion of a small field limiting aperture that cuts away electrons emitted outside the field of view reduces the electron beam intensity behind that point and improves the lateral resolution. This, together with accurate photon flux reduction, led to a demonstrated lateral resolution of 18 nm in XPEEM [62]. This limit is still far from analogue result in LEEM, but there are reasons to think that a future, dedicated instrument could bridge the gap, at least partially. In particular, such instrument should be equipped with apertures at the backfocal plane in order to cut non-necessary electrons with different emission angle, and at the first dispersive plane inside the beam separator, to roughly filter away secondary electrons. High efficient detectors and improved illumination optics would also help to obtain good quality images with lower photon flux. These measures to reduce the space charge effect will result in higher resolution.

## 4.1 Imaging mode

### 4.1.1 PEEM and XPEEM

The imaging mode with photoemitted electrons displays a magnified real-space image of the probed surface. Such image contains valuable information on the physical and chemical state of the specimen. The contrast mechanism in PEEM has been already discussed: in the case of low energy photons, areas can emit a different number of electrons as a function of the local work function and of the local density of states of the valence band. In the case of energy-filtered XPEEM with core level electrons, the amplitude contrast can be also given, in first approximation, by the relative local abundance of the particular element. The photon energy is often selected in order to maximize the image intensity, in agreement with the atomic cross section of the selected emission line and the brilliance of the photon source. The resulting kinetic energy of the photoemitted electrons is typically ranging between 50 and 200 eV, so that the small probing depth due to the inelastic mean free path influences the contrast. **Figure 16** shows a direct example of XPEEM images with chemical contrast. In this case, an Ir(100) surface is partially covered by single and multilayer graphene [126]. XPEEM images using Ir $4f_{7/2}$ photoemitted electrons (**Fig. 16a**) shows areas with different intensity. The attenuation of the signal is due to the screening from graphene layers of different thickness. This is confirmed by XPEEM images taken with C 1s electrons (**Fig. 16b**), where the intensity is proportional to the graphene thickness.

The screening effect can be evaluated quantitatively by changing the start voltage and extracting the IV curve, with the same procedure seen for LEEM. In this case, the collected intensity curve is the local spectroscopic photoemission peak, i.e. a stack of XPEEM images gives access to local XPS spectra. **Figure 16c** and **d** display, respectively, the Ir 4f and C 1s core level emission spectra, obtained by integration over selected regions of interest over a stack of XPEEM images at different start voltages. The intensity drop in the Ir 4f photoemission line follows the exponential damping factor $e^{-d/\lambda_{gr}(E_0)}$, where $d$ is the layer separation and $\lambda_{gr}(E_0)$ is the kinetic energy dependent effective attenuation length [127]. The intensity growth of C 1s core level shows a similar behavior, although it does not increase linearly with the thickness. This deviation is given by the screen effect of the outermost graphene layers that diffract the photoelectrons emitted from buried C atoms. This example demonstrates how the intensity evaluation of XPEEM imaging provides an alternative method of assigning the thin film thickness.

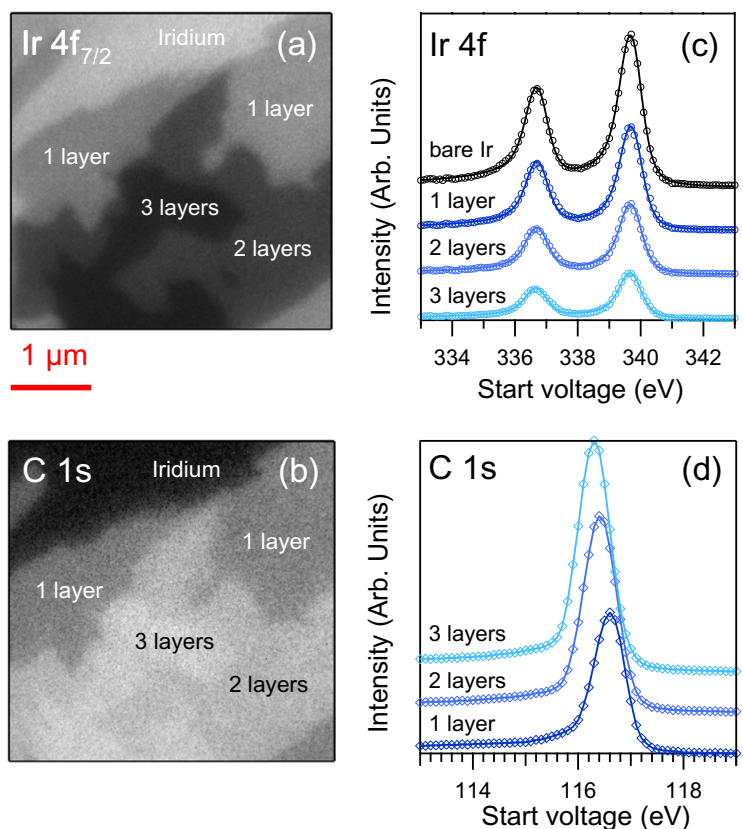

**Figure 16:** (a) XPEEM Ir 4f image of a multi-thickness graphene island. The bare iridium surface is visible on the top; (b) XPEEM C 1s image of the same surface region; (c) Ir 4f and (d) C 1s XPS spectra measured in the regions indicated by the labels. The start voltage corresponds to the kinetic energy of the photoemitted electrons. A photon energy of 400 eV was used in the experiment. Reproduced from Ref. [126] with permission, copyright 2014 Elsevier.

## 4.1.2 Brightfield and Darkfield PEEM

The ordinary microscope setup in PEEM includes the insertion of a contrast aperture to limit the acceptance angle in order to reduce spherical aberration and enhance contrast and resolution. Usually, the aperture selects only electrons with normal emission, i.e. limits the diffraction pattern to a neighborhood of the Γ point of the first Brillouin zone (FBZ). As in the case of LEEM, it would be convenient to build PEEM images with electrons coming from other regions of the first Brillouin zone [128, 129]. The methods for implementing darkfield measurement has already been reviewed in Sect. 3.1.2 and schematized in **Fig. 8**. In case of emitted electrons, only the first two are available, i.e. (i) to position the contrast aperture around the desired *k*-point of the reciprocal space and (ii) to incline the sample tilt by a certain angle. In addition, it is possible to deflect the emitted beam after the objective lens. Among the three, the second offers the best advantages in terms of PEEM image quality, since it is the one that let the electrons travel along the optical axis. It should be remarked that the required tilt inclination is reduced by the change of coordinates due to the accelerating field of the cathode immersion lens. It is sufficient to incline the tilt angle by less than 2° to obtain a consistent shift of the reciprocal image in the backfocal plane of the objective lens and to get to the edge of the first Brillouin zone of most materials. The best procedure to select the desired diffraction feature for imaging is to display the backfocal plane and to modify the sample tilt mechanically until the selected point is positioned at the center. Once the contrast aperture is placed, the microscope can be shifted to imaging mode, whereas the intensity is proportional to the local density of states at the selected *k*-point.

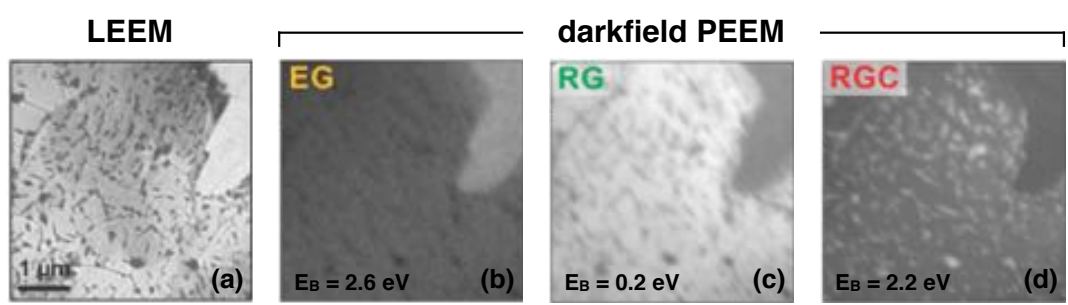

Figure 17: (a) LEEM and darkfield PEEM of epitaxial-EG (b) and rotated-RG (c) graphene flakes grown on Ni(111). RGC patches (d) are non-interacting graphene on a Ni-carbide interstitial layer. Reproduced from Ref. [130].

Darkfield PEEM is very powerful to probe the local electronic properties of heterogeneous surfaces. For example, it has been employed on graphene grown on Ni(111), a model case for the study of interfacial interactions that can play a key role in tailoring the magnetic [131] and chemical [132] properties of 2D materials. In particular, darkfield PEEM was used to determine the lateral extension of epitaxial and rotated monolayer graphene grown on Ni(111), and to identify particular areas in which the rotated graphene is metallic due to the formation of an interstitial carbide layer [130].

Brightfield LEEM image of the system (**Fig. 17a**) reveals the presence of three coexisting phases with different grey levels. Characteristic IV reflectivity curve and µ-LEED measurements (not shown here) assigned the brighter area in the upper right side to epitaxial graphene (labeled EG), while the neutral gray area and the dark grey patches are rotated graphene (labeled RG). Local XPS measurement on the C 1s core level performed with the method addressed in the previous Section proved that the dark patches are small areas in which the monolayer graphene rests upon a Ni-carbidic interface layer (labeled RGC) [133]. Darkfield PEEM was then performed with electrons photoemitted from graphene's $\pi$–band. When the reciprocal space $K$ point of the epitaxial graphene is selected, only the upper right area becomes bright (**Fig. 17b**). When the equivalent point of the rotated graphene is used, the contrast shows a dependence on the selected start voltage. The RGC phase exhibits high density of states at a binding energy of 0.2 eV (**Fig. 17c**), while the RG phase shows the same behavior only at 2.2 eV (**Fig. 17d**) [134]. This demonstrated that the $\pi$–band of RGC graphene has the Dirac point very close to the Fermi energy, i.e. shows an almost metallic state, while in EG and RG it undergoes a severe hybridization with the substrate states. This result, together with the angular-resolved photoelectron spectroscopy measurement, demonstrated that the presence of the interstitial carbide layer decoupled the monolayer graphene from the substrate.

### 4.1.3 XAS-PEEM and Magnetic PEEM Imaging

It has been shown that when X-rays illuminate a sample, the displayed intensity of slow secondary electrons exceeds by orders of magnitude the one of photoemitted electrons. It is therefore convenient to use them for imaging, enabling even video rate acquisition. Moreover, imaging with secondary electrons is particularly useful in presence of tunable X-ray sources such as synchrotron beamlines. Direct secondary PEEM imaging during photon energy scan reveals intensity variation due to the X-ray absorption process, i.e. the characteristic curve for a selected region of interest corresponds to the local X-ray absorption spectrum (XAS) [135, 136]. Compared to XPS, the XAS has a bigger signal-to-noise ratio and uses electrons with a larger IMFP, thus enabling a larger probing depth. Moreover, it can be performed even without active energy filtering, as the photoemitted electrons give a negligible contribution to the overall intensity. The possibility to select a desired photon polarization gives also access to magnetic characterization of materials [102]. In the case of circular polarized light, one can use the X-ray Magnetic Circular Dichroism (XMCD) to obtain magnetic contrast in PEEM. In XMCD the angular momentum of circular polarized light is used to impose selection rules over the photon absorption process [137]. In ferromagnetic materials, the valence and the conduction band are split in two electron populations, a majority with spin parallel to the magnetization vector and a minority with spin antiparallel. In the photon absorption process, the Fermi's Golden Rule imposes that the transition probability of an electron from the initial core level to an unoccupied state depends on their density of states. Therefore, the different final density of unoccupied states in the majority and minority population determines two distinct absorption probabilities for left and right circular polarized light. The magnetic dichroism is defined by convention as the intensity difference between the two absorption spectra, $I_{XMCD} = I_{1\downarrow} - I_{1\uparrow}$ and is directly proportional to the scalar product between the magnetic moment of the sample and the photon polarization vector. XMCD can be performed in PEEM by taking two images with opposed circular polarization at the absorption photon energy and subtracting each other, thus extracting the exchange asymmetry image, as in the case of SPLEEM,

$$I_{XMCD} = \frac{I_{1\downarrow} - I_{1\uparrow}}{I_{1\downarrow} + I_{1\uparrow}} \ .$$

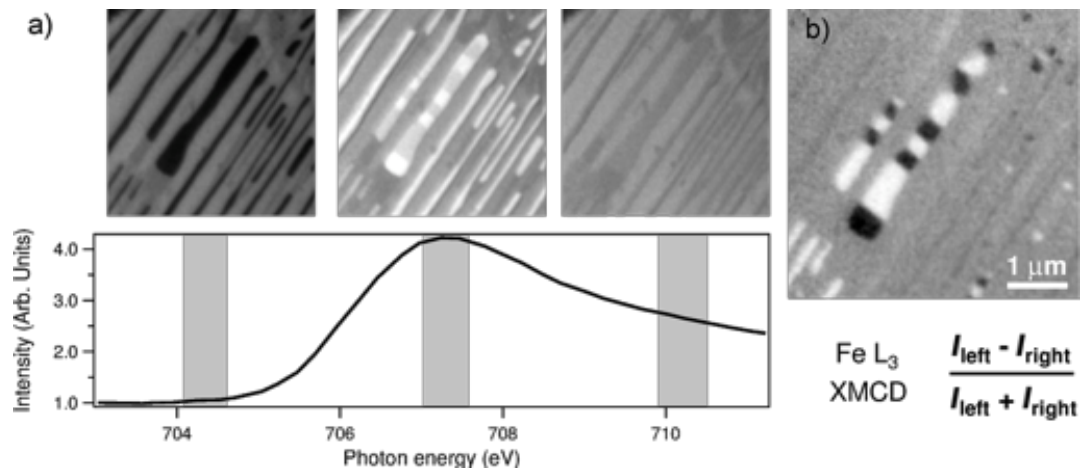

Figure 18: (a) illustration of imaging spectroscopy in XAS mode. Fe nanowires on W(110) appear dark on the left panel at a photon energy of 704.5 eV. At the Fe $L_3$ threshold, the wires become much brighter (middle panel). The XAS spectrum below is extracted from the largest nanowire in the center. (b) Illustration of XMCD-PEEM imaging. The photon energy is tuned to the $L_3$ maximum. The field of view is 5 μm. The start voltage is 3 eV in order to collect secondary electrons. Within the image plane, the X-ray direction is perpendicular to the nanowire axis. Reproduced from Ref. [44].

The XAS-PEEM imaging process is illustrated in **Fig. 18a**, where Fe nanowires on a W(110) surface are displayed [44]. When the photon energy is set to be off-resonance (left), the contrast is given simply by the different secondary yield. Once the photon energy is set to the Fe absorption $L_3$ edge (center), the nanowires become brighter whereas the clean W(110) surface in between shows no change. The contrast reduces again for higher, above-resonance photon energies (right), although it is changed also as the new Fe $2p_{3/2}$ photoemission channel enhances the Fe secondary yield. The entire absorption spectrum of an individual nanowire (below) can be extracted from local integration over a stack of PEEM images at different photon energy. Then, XMCD imaging is obtained by tuning the photon energy to the Fe $L_3$ adsorption edge and taking the difference between two secondary PEEM images with opposed circular polarization (**Fig. 18b**). The wires with magnetization perpendicular to the polarization direction appear grey as the scalar product reduces the XMCD signal to zero. The black and white regions in the central wires are dipolar domains with magnetization parallel and antiparallel to the beam direction, respectively. Note that the strong magnetic contrast in the case of highly magnetic materials is visible even at the single

XAS image acquired with one circular polarization. The most advanced PEEM systems are then capable of acquiring magnetic images at video rate, thus enabling dynamic studies of the magnetic domain evolution in non-equilibrium conditions.

In particular samples, magnetic imaging can be performed also using X-Ray Magnetic Linear Dichroism (XMLD). The sum rules are similar to those for XMCD [102], so the technical implementation mimics the one already discussed. Its main application field is the study of antiferromagnetic materials and their interfaces with ferromagnetic materials, as well as multiferroics. In conclusion, it should be noticed that in a synchrotron beamline equipped with insertion devices, the polarization vector could be varied only in the plane perpendicular to the photon beam propagation direction. Therefore, XMCD-XMLD contrast can be achieved only for selected geometries. PEEM systems with perpendicular illumination can characterize only out-of-plane magnetization states, while grazing photon incidence gives its best with in-plane states.

## 4.2 Diffraction mode

### 4.2.1 µ-ARPES and µ-XPD

As in the case of µ-LEED, the diffraction operation mode of a PEEM system gives access to the angular distribution of energy-filtered photoemitted electrons. The convenient placement of a field limiting aperture on an intermediate image plane selects a probed region down to 1 µm in size. The Angle-Resolved Photoelectron Spectroscopy (ARPES) is one of the most important methods to study the band structure of solids [138, 139]. The direct display of the entire angular spectrum in a single image gives PEEM systems the possibility to acquire ARPES spectra over a large energy range in a very short amount of time without any mechanical movement of the sample. The same operation mode is then capable of detecting X-ray Photoelectron Diffraction (XPD) spectra, when the angular distribution of photoemitted core level electrons is displayed. Although the energy resolution is not comparable to a dedicated ARPES machine, its fast acquisition and the possibility to conjugate it with microscopy measurement makes PEEM systems ideal for studies on dynamic and/or heterogeneous systems [140].

ARPES spectra are usually collected at relatively low photon energy. Typical photon sources are lasers under high harmonic generation (the fourth harmonic of a Ti:sapphire laser delivers 6 eV photons), He discharging lamps (HeI emission line at 21.2 eV) and synchrotrons (10-100 eV). The latter is more indicated to highlight surface states, due to the smaller IMFP of photoemitted electrons. Among others, µ-ARPES has been employed to access the $\pi$-band of graphene and to quantify the local doping induced by foreign species and/or interaction with the substrate [141]. In particular, the local probing in diffraction mode was crucial to determine the doping level of 2D-heterojunctions created with selected-area low-energy irradiation of a monolayer graphene with $N_2^+$ ions at room temperature [142]. **Figure 19a** shows a LEEM image of the boundary region between irradiated and non-irradiated graphene grown on Ir(111) surface. The first appears notably darker than the second due to the development of corrugations and defects. Microprobe ARPES measurements were performed separately on the two areas (**Fig. 19b**). The reciprocal space image, taken with photon energy of 40.5 eV and at a binding energy of 0.45 eV, displays a large density of states around the K–point due to the graphene band structure, while the background modulation resembles the surface states of Ir(111) substrate. While in the non-irradiated area the circular feature around the K–point corresponds to a slice of the Dirac cone, the irradiated area showed no band splitting at this point. The entire band structure can be depicted if one extracts the intensity line profile along a high-symmetry direction over a stack of ARPES images at different kinetic energy. **Figure 19c** shows the Momentum Distribution Curve (MDC) along the normal to Γ-K direction on the K–point for both areas. The non-irradiated graphene presented a linear dispersion of the $\pi$-band centered at the Dirac point $E_D = 0.08 \pm 0.03$ eV, corresponding to a slight *p*-doping due to the interaction with the substrate. The irradiated graphene showed a large shift of the Dirac point towards higher binding energies, quantified as $E_D = -0.45 \pm 0.07$ eV. This *n*-doping was induced by the implantation of N atoms in the C mesh of graphene [143]. This example shows clearly how microprobe ARPES in PEEM systems has the power to characterize the electronic states of micron-sized regions.

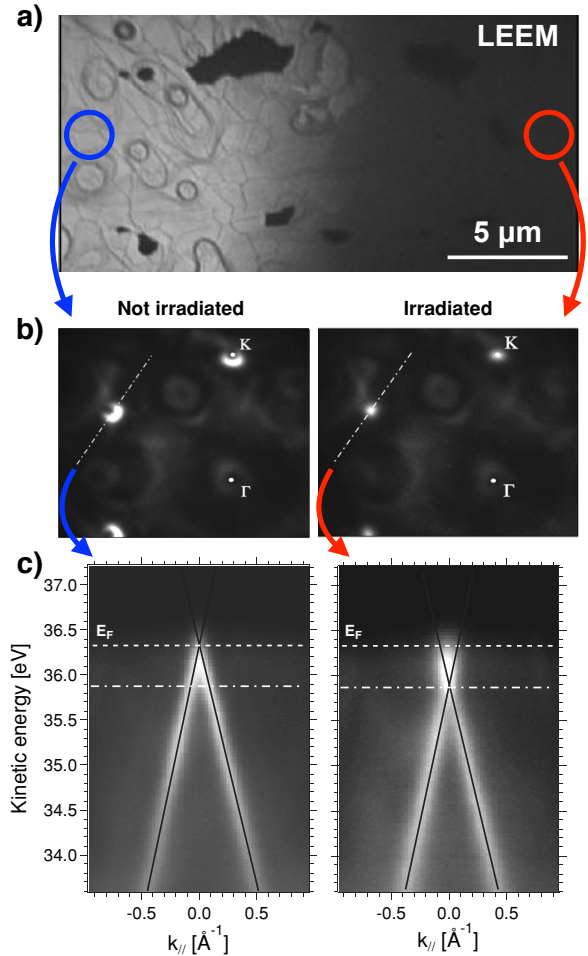

Figure 19: (a) LEEM image ($E_0$ = 7 eV) of the boundary between $N_2^+$ irradiated (dark) and non-irradiated graphene (bright) grown on Ir(111). (b) µ-ARPES patterns of non-irradiated (left) and irradiated (right) epitaxial graphene on Ir(111) before annealing. The image was acquired 0.45 eV below the Fermi level. The dash-dot line, passing through the K point identifies the high symmetry direction perpendicular to ΓK; hν = 40.5 eV. (c) Intensity cut in the plane passing through K along ΓM. The MDCs demonstrate linear dispersion of the π-band close to the Fermi energy $E_F$ (left). On the right, the same intensity cut for an $N_2^+$ irradiated sample (3 min at $p_{N2}$ = 2×10$^{-5}$ mbar, ion current 0.14 µA). The Dirac energy was estimated by fitting the MDC and determining the intersection of the resulting solid lines. $E_D$ = −0.45 ± 0.07 eV. The Fermi energy $E_F$ is highlighted by the dashed line, while the dash-dot line represents the energy of the images in (b). Reproduced from Ref. [142] with permission, copyright 2015 Wiley-VCH Verlag.

The SPELEEM microscope used in these measurements is an energy-filtered LEEM/PEEM station dedicated to the probe of surfaces with backscattered electrons and synchrotron radiation. The lens setup is designed to guarantee good performances overall, but is not devised for state-of-the-art measurement in every single operational mode. The provided energy resolution, 330 meV, was adequate to detect accurately the shift of the graphene $\pi$-band, but was insufficient to determine whether a small bandgap has been opened by the substitutional implantation of N [141]. PEEM stations specifically designed for high performance in ARPES, e.g., NanoESCA [45, 140], forgoes the lateral resolution for a better energy and angular resolution. Such systems are often labeled as momentum microscopes or $k$–PEEM.

## 4.3 Spectroscopy mode

### 4.3.1 μ-XPS

The spectroscopy analysis of photoemitted electrons is a major tool in surface science. Since its first development by Kai Siegbahn in 1957, the Electron Spectroscopy for Chemical Analysis showed how the accurate measurement of photoemitted core level electrons allows the retrieval of information on the chemical structure and bonding, the elemental composition and the surface state of atoms [144]. PEEM systems equipped with an energy analyzer are capable to perform PhotoEmission Spectroscopy (PES) from a selected area of the sample, much smaller in size than a standard PES machine [145]. The probed area can be selected by strong demagnification of the illuminating photon beam and insertion of a field limiting aperture in the first intermediate image plane. The best PEEM systems available worldwide for PES measurements make use of synchrotron radiation and take advantage of its high brilliance and energy tunablity. The access to the dispersive plane in PEEM enables direct measurement of the photoemission spectrum from a localized area and is often called microprobe X-ray Photoelectron Spectroscopy (μ-XPS).

The standard setup for a PEEM in dispersive mode includes the insertion of the field limiting aperture at the intermediate image plane and the contrast aperture at the diffraction plane (**Figure 3**), limiting conveniently the probed area and the acceptance angle. The dispersive plane is then projected on the detector and appears as a thin line with modulated intensity. Electrons with initial kinetic energy equal to the bias value $eV_0$ are projected at the center. The width of the energy window depends on the final magnification of the dispersive plane and the construction constraints, and can be as wide as ±5 eV. The energy spectrum around the bias value is then revealed with an intensity line profile after opportune background subtraction and response function division. The photoemission spectrum collected in dispersive mode has a better energy resolution and a more favorable signal-to-noise ratio than the one extracted from a stack of XPEEM images over a start voltage range. In several cases the signal-to-noise ratio is sufficiently high to enable fast image acquisition up to video rate. This open the gates to the microprobing of dynamic processes in real time, such as catalytic reactions, thin film growth, deposition of atoms and molecules on surfaces and testing in changing conditions of temperature and pressure. The data quality is well visible in the XPS data measured for the heterogeneous system already presented in Sect. 4.2.1. In **Fig. 20a** and **b** N 1s and C 1s core level photoemission spectra are depicted, respectively, of a monolayer graphene grown on Ir(111) and locally irradiated with low-energy $N_2^+$ ions at room temperature [142]. It has been already shown with μ-ARPES that the ion irradiation induces a negative doping on the graphene sheet. XPS N 1s spectrum of the irradiated graphene describes in great detail the film stoichiometry and the atomic arrangement of

N atoms in the C mesh. The photoemission peak can be fitted with four Doniach–Šunjic line profiles [141, 143, 146], later assigned to pyridine-like, twofold coordinated nitrogen (N1), pyrrole-like, threefold coordinated nitrogen (N2) and substitutional nitrogen (N3–graphitic and N4–secondary graphitic) on the base of their binding energy. By evaluating the areas of the various components, one can establish the relative abundance of the defects and obtain that 43% of the N is in substitutional configuration. N 1s spectrum of non-irradiated graphene confirms that no N atoms are detected few µm aside. The analysis of C 1s spectra shows a notable broadening for the irradiated area at higher binding energies with respect to the main emission line (the only component in the non-irradiated area). This is consistent with the picture of graphene functionalized with N. The irradiated spectrum was fitted considering the different contributions of $sp^2$ and $sp^3$ carbon, vacancies and carbon species bonded with nitrogen [147, 148]. The evaluation of peak areas allows to estimate a vacancy abundance of 1.7% of a ML, evidencing the little damage suffered by graphene. Moreover, from the comparison of N 1s and C 1s peak areas one can estimate the overall N coverage. Since both core levels have the same number of electrons and were measured at the same photon energy ($h\nu = 500$ eV), the N/C peak area $A$ ratio can be normalized with the Yeh and Lindau photoionization cross section $\sigma$ [149] and the energy-dependent transmission of the microscope, which in the case of SPELEEM is proportional to the kinetic energy $E_0$ of the photoemitted electrons. In case of coplanar species, no IMFP correction is needed (as in Sect. 4.1.1), therefore the nitrogen abundance is equal to

$$\theta_N = \frac{A_{N1s}}{A_{C1s}} \frac{\sigma_{C1s}}{\sigma_{N1s}} \frac{E_0^{N1s}}{E_0^{C1s}}$$

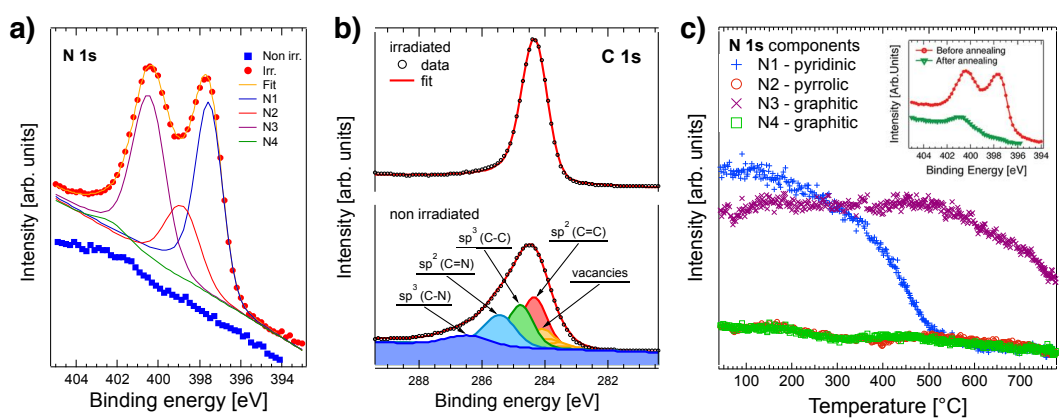

**Figure 20:** (a) N 1s spectra and (b) C 1s spectra obtained from the non-irradiated and $N_2^+$ irradiated graphene on Ir(111). The best fit to the experimental data is also shown: the simulated curve for the irradiated graphene is shown as the sum of different components, each with a Doniach–Šunjic line profile. (c) Evolution of the intensity of the N 1s components, N1–N4, plotted as a function of the annealing temperature. In the inset the N 1s spectra of ion irradiated graphene before and after annealing at 800 °C are compared; hν = 500 eV. Reproduced from Ref. [142] with permission, copyright 2014 Wiley-VCH Verlag.

The stability test of locally implanted nitrogen took full advantage of the real time capability of µ-XPS. The annealing in UHV up to a temperature of 800 °C was followed probing the N 1s photoemission line every 2 s. The intensity of the various components, as determined in a fit, is presented in **Fig. 20c**. As can be seen, both N2 and N4 components decrease steadily with increasing temperature, reaching negligible intensity at high temperature. The N1 component, associated with intercalated, pyridine-like nitrogen, decreases moderately up to 430 °C and suddenly decays above this temperature. The N3 component, related to graphitic nitrogen, remains constant up to 500 °C and decays moderately with annealing temperature. After annealing, the overall nitrogen coverage decreased from 0.07 ML to 0.04 ML. The real time µ-XPS highlighted the different evolution of each defect species during annealing, giving highly valuable information on the stability of doping level for annealed heterostructures embedded in a single graphene sheet.

The dispersion mode can probe photoemitted electrons at any given start voltage. Therefore, it is suitable to probe not only core level electrons, but also valence band electrons and even secondary electrons. It should be noticed that the size of the contrast aperture determines the angular acceptance, i.e. the probed region in the reciprocal space. Usually the neighborhood of $\Gamma$ in the FBZ is selected, but one can use the methods introduced for darkfield measurement (Sect. 4.1.2) to filter another region of the $k$-space. Very often, core level XPS measurement is combined with valence band measurement up to the Fermi edge, to refer the binding energy. The Fermi edge of conductive surfaces at low temperature is frequently used to test on the go the energy resolution of the instrument. The edge of the secondary yield can be also probed to measure the work function differences among separate areas.

## 5. Perspectives

More than 80 years after the first PEEM image, cathode lens microscopy continues to be one of the most used techniques in surface science. The ability to conjugate microscopy, spectroscopy and diffraction in a single instrument is the key of its successful application in several fields. The increasingly high level of sophistication led to the construction of systems with many specializations, such as magnetic imaging, pump-probe photoemission, synchrotron endstations, momentum microscopes and more. The growing user community shares its findings and perspectives in many dedicated workshops; among others, the LEEM-PEEM biennial gathering reviews the status of LEEM, PEEM, SPLEEM, XPEEM and related techniques, promotes applications of cathode lens microscopy to a broader audience and highlights the most recent scientific advances and instrumental developments.

After the advent of synchrotron radiation, spin polarized electron guns and pulsed laser in the attosecond domain, cathode lens microscopy can be defined as a mature and well established technique. In the more recent years, much of the technical effort is spent to address three main problems, here reviewed briefly:

**Reduction of space charge effect**: its inevitability imposes the creation of a system expressly designed to circumvent it. Such apparatus must have particular filters in the backfocal plane, in the first intermediate image plane and in the first dispersive plane, all of them placed very close to the objective lens. Moreover, it must be equipped with state-of-the-art detection system. The new generation of SMART microscope, called SMART 2, is currently (2017) under construction and implements most of the strategies discussed to give full advantage of aberration correction in XPEEM. This development is needed for advance in pump-probe experiments.

**Resolution improvement**: in principle, the aberration correction via electrostatic mirror pushes LEEM towards the ultimate resolution, i.e. the diffraction limit. Several groups, with remarkable mention of Ruud Tromp's team, have investigated theoretical and experimental aspects of aberration correction. They discovered that the correct state is intrinsically unstable due to the performance of electronics. The creation of more stable devices and a dynamic feedback system that prolongs the lifetime of the correct state remains a serious challenge.

**Pressure gap**: in a standard cathode lens microscope, the high electrostatic field of the accelerating stage limits the pressure in the main chamber to be lower than ~ $10^{-5}$ mbar. The direct observation of surfaces under near-ambient pressures is highly desirable, e.g., in the field of catalysis. In principle, the implementation of high-pressure cathode lens microscopy can take profit of the technical development of near-ambient pressure XPS systems, already available since late 2000s. These systems are equipped with a pressure cell that maintains high pressure in a small volume in front of the sample and let photoemitted electrons escape through nozzles and membranes towards the detector placed in UHV condition. The design of a high-pressure cell that works with the strong electrostatic fields required in cathode lens microscopy is a very difficult task that may require years of research and development. A successful breakthrough in the pressure problem can breathe new life into the cathode lens microscopy user community and ensure a bright future for this technique.